\documentclass[useAMS,usenatbib]{mnras}
\usepackage{natbib}
\usepackage{graphicx}

\title[SPS-IGIMF code]{Stellar population synthesis models with a physically varying IMF}

\author[Zonoozi et al.]
{Akram Hasani Zonoozi$^{1,2}$\thanks{
E-mail:  \mbox{a.hasani@iasbs.ac.ir} (AHZ)}, Hosein Haghi$^{1,2,3}$
and  Pavel Kroupa$^{1, 4}$,
\\
$^{1}$Helmholtz-Institut f\"ur Strahlen-und Kernphysik (HISKP), Universit\"at Bonn, Nussallee 14-16, D-53115 Bonn, Germany\\
$^{2}$Department of Physics, Institute for Advanced Studies in Basic Sciences (IASBS), PO Box 11365-9161, Zanjan, Iran\\
$^{3}$School of Astronomy, Institute for Research in Fundamental Sciences (IPM), PO Box 19395 - 5531, Tehran, Iran\\
$^{4}$Charles University in Prague, Faculty of Mathematics and Physics, Astronomical Institute, V Hole\v{s}ovi\v{c}k\'ach 2, CZ-180 00 \\
Praha 8, Czech Republic\\}

\begin{document}

\date{Accepted .... Received}

\pagerange{\pageref{firstpage}--\pageref{lastpage}} \pubyear{2013}

\maketitle

\label{firstpage}

\maketitle

\begin{abstract}

Interpreting galactic luminosity requires assumptions about the galaxy-wide initial mass function (gwIMF), often assumed invariant in most stellar population synthesis (SPS) models. If stars form in clusters with metallicity- and density-dependent \textit{stellar IMFs}, the integrated galaxy-wide IMF (IGIMF) can be calculated, with its shape depending on the star formation rate (SFR) and metallicity. The shape of the IGIMF thus depends on the star formation rate (SFR) and metallicity. We develop the \texttt{SPS-VarIMF} code which enables us for the first time to compute the spectra, luminosities, and remnant populations of galaxies in the context of the varying gwIMF with time, SFR, and an assumed metallicity. Using the \texttt{SPS-VarIMF} code one can calculate how the interpretation from the integrated galactic light may change if the underlying galaxy-wide IMF is assumed to be environmentally dependent instead of being invariant. In particular,  we compare the time evolution of the galaxy color and the stellar mass-to-light ratio in different bands for the IGIMF  and invariant canonical gwIMF assuming constant and delayed-$\tau$ star formation histories. We show that the underlying gwIMF can be determined by examining the colors and luminosities of late-type galaxies in UV and optical bands. On the other hand, for early-type galaxies, it is difficult to distinguish which gwIMF is valid since adopting the different gwIMFs yields almost identical colors. However, their gwIMF-dependent $M/L$ ratios differ by up to an order of magnitude. Massive present-day elliptical galaxies would have been $10^4$ times as bright as at present when they were forming.

\end{abstract}


\section{INTODUCTION}\label{Sec:INTRO}


The initial stellar mass function\footnote{Hereinforth the stellar IMF constitutes the distribution of all initial stellar masses in an embedded star cluster which forms in a molecular cloud clump, ie. an embedded star cluster. The gwIMF is the galaxy-wide distribution of initial stellar masses and is thus the sum of all stellar IMFs in a galaxy.}  (IMF), which describes the mass distribution of newly formed stars in a molecular cloud clump, is one of the most fundamental functions in astrophysics and plays an important role in galaxy evolution \citep{Kroupa13, Hopkins18, Kroupa2024}. The IMF is needed to model the chemical enrichment and population synthesis of galaxies \Citep{Tinsley80} and to estimate their stellar mass-to-light ($M/L$) ratios and baryonic mass content from the observed luminosities. It is also an important key for understanding the formation of star clusters and their dynamical evolution.

The direct observational determination of the stellar IMF is extremely challenging as it is based on star counts and remnant-star corrections and it evolves in time. Much work has been done to constrain the shape of the IMF on star-cluster and galaxy scales. Many biases, assumptions, and systematic uncertainties are involved in the derivation of the IMF, i.e., measuring the luminosity function from the observational data, converting it to the present-day MF (PDMF) using a correct stellar mass-luminosity relation, and correcting the PDMF for the star formation history (SFH), stellar evolution, binary fraction and dynamical evolution of star clusters \citep{Scalo86, Kroupa93, Kroupa01, Kroupa02, Bastian10, vanDokkum11, Kroupa03, Hopkins13, Matteucci90, Vazdekis03, Hoversten08, Meurer09, Lee09, Gunawardhana11,  Bastian10, Kroupa13, Offner14, Kroupa2024}. 

The universality or environmental dependency of the stellar IMF is one of the important questions in modern astrophysics.  While it is common to assume that the stellar IMF is invariant, recent observations suggest that the IMF varies in different environments \citep{Kroupa02, Gunawardhana11, Dabringhausen12, Marks12}. 
The low-metallicity starburst region 30 Doradus in the LMC \citep{Schneider18, Banerjee2012}, the low-metallicity massive cluster NGC 796 in the LMC/SMC bridge region \citep{Kalari18}, and the low-metallicity star-forming region in the outer-Galaxy Sh 2-209 \citep{Yasui2023, Zinnkann2024} have been found to have top-heavy IMFs for stars above $1 M_{\odot}$.  

Extragalactic observational data show a systematic variation of the galaxy-wide IMF (gwIMF). The gwIMF seems to be more top-heavy for active galaxies with high star formation rates (SFRs) \Citep{ Hoversten08, Lee09, Meurer09, Habergham10, Gunawardhana11}, while dwarf galaxies and low surface brightness galaxies seem to have top-light gwIMFs \Citep{Ubeda07, Lee09, Watts18}. 
\Citet{Zhang18} found evidence of a top-heavy gwIMF in a sample of starburst galaxies at redshifts of around two to three.  A varying gwIMF at the low mass end has been found in early-type galaxies \Citep{vanDokkum11, Spiniello15, Conroy17} and in ultra-faint dwarf galaxies, being more bottom-heavy for more massive galaxies and more bottom-light\footnote{Here "top" refers to stars above $1 M_{\odot}$ and the "bottom" means $m \leq 0.5 M_{\odot}$. "Top heavy" thus means that the IMF has a flatter shape above $1 M_{\odot}$ compared to the canonical stellar IMF, while "top light" gwIMF or IMF has a steeper declining slope therewith lacking massive stars. Similarly for "bottom heavy" (more) and "bottom light" (lacking low-mass stars).} with decreasing galactic mass and metallicity, respectively  \Citep{Geha13, Yan2017, Gennaro18, Dabring2023, Yan2024}.

The modern integrated galactic IMF (IGIMF)  theory of the gwIMF is defined based on optimal sampling \Citep{Kroupa13, Yan2017} instead of the stochastic star formation scenario as used originally by \citet{Kroupa03} to calculate the gwIMF, by adding up the stellar IMFs of all freshly formed embedded clusters within a galaxy. According to this theory, the gwIMF may significantly deviate from the star-cluster-scale stellar IMF. The IGIMF theory suggests that the gwIMF in massive galaxies with high SFRs is top-heavy and bottom-heavy at large metallicity, while in low-mass galaxies with low star formation activity, it becomes top-light \citep{Yan2017}. A complete grid of the IGIMF is available for galaxies of various SFR and metallicity, as provided by \citet{Jerabkova2018} and \citet{Haslbauer2024}. Using the IGIMF theory, the chemical evolution of the solar neighborhood \Citep{Calura10},  and of star-forming and elliptical galaxies \Citep{Kopen2007, Recchi09, Fontanot2017, Yan2017, Yan2020, Yan2021} has been studied. Many other aspects of galaxy evolution are resolved using the IGIMF theory (for a review see \citealt{Pflamm11}), and for example the radial $H\alpha$ cutoff in disk galaxies with UV extended disks is a normal consequence of the IGIMF theory \citep{Pflamm08}, as is the existence of the galaxy mass-metallicity relation \citep{Kopen2007, Recchi2015, Yan2021, Haslbauer2024}.


The stellar $M/L_X$ ratio of a galaxy in some photometric passband $X$ strongly depends on the gwIMF and the SFH of the galaxy and links photometric observations to dynamics.  Determining the stellar mass of galaxies based on the amount of light they emit is necessary to study the dynamics and evolution of galaxies. In spiral galaxies, the rotation curve is sensitive to the gravity produced by the total mass of the galaxy. Decomposing the rotation curves into the contributions of gaseous, stellar, and (phantom\footnote{"Phantom" stands either for the yet-to-be-confirmed dark matter in the $\Lambda$CDM cosmological model, or the purely mathematical source of the Milgromian potential in MOND. }) dark matter contents therefore allow us to constrain the structure of the (phantom) dark matter and alternative theories of gravity such as Modified Newtonian Dynamics (MOND, \citealt{Milgrom1983, Famaey2012, Banik2022, Kroupa2023b}). While the gas contribution is well understood from the  21 cm HI emission, a challenge is to gain knowledge of the stellar mass contribution  (including the remnants) for which the value of the $M/L_X$ ratio must be determined.

The SFHs of galaxies reveal how galaxies evolved to what we see today. Newborn stars produce new light and change the chemical composition of the interstellar gas through stellar winds and supernova explosions. This causes the next stellar population to form with a new metallicity. Assuming that the gwIMF is invariant, the $M/L_X$ value does not depend on the overall normalization factor of the SFR, since both the mass and light are integrals over the SFH. Therefore, the $M/L_X$ ratio resulting from an invariant gwIMF will be entirely independent of the total mass of the galaxy. The situation is more complicated if the gwIMF changes with different star formation conditions. Adopting the IGIMF theory, the gwIMF varies systematically with the galaxy’s SFR and metallicity. This means that in the IGIMF context, not only the functional form of the SFR but also the total mass of the galaxy plays an important role in its present-day $M/L_X$ ratio. The SFR, which determines the galaxy's total stellar mass, is important because it affects the shape of the gwIMF and the fraction of high-mass and low-mass stars that form at a given time in a galaxy \citep{Jerabkova2018, Haslbauer2024}.

Stellar population synthesis (SPS) is fundamentally important for determining the stellar content in external galaxies where individual stars cannot be resolved \citep{Tinsley80, Bruzual03, Vazdekis2012}. The gwIMF and SFH define the stellar population, determine the masses of living stars and remnants, and impact the matter cycle and chemical enrichment. Several groups have based their SPS models on an invariant gwIMF at a specific metallicity and using standard stellar evolutionary tracks \citep{Bruzual03, Conroy2009}.

\citet{Stanway2019} and \citet{Stanway2020} explored the impact of varying the shape of the IMF and resampling uncertainties on the initial assumptions regarding the mass dependence of the binary fraction, mass ratio, and period distribution, on the observable properties of a binary stellar population synthesis model. The initial parameter distributions for the synthetic populations were sampled stochastically. In addition,  the Binary Population and Spectral Synthesis (BPASS code) includes binary evolution in modelling the stellar populations \citep{Eldridge2017, Eldridge2022}. A recent paper \citep{Stanway2023} evaluates the relative impact of the most important sources of uncertainty in binary stellar populations, constructed from a large grid of detailed binary stellar evolution models focusing on the effects of stochastic sampling on uncertainties in the integrated light of the population.

Deducing $M/L_X$ ratios in star-forming galaxies, such as spirals and irregular galaxies, is challenging due to the conflicting effects of star formation and chemical enrichment \citep{Bell2001}. However, extracting the  $M/L_X$ ratios of quiescent galaxies like ellipticals seems to be straightforward in the context of the invariant gwIMF \citep{Schombert2016}.  In this paper, we investigate the importance of considering the gwIMF variation on the SPS models and our interpretation of the light, color, and spectra we get from a galaxy.  We aim to investigate how  $M/L_X$ varies among galaxies of different masses, metallicities, and SFHs using the invariant canonical IMF as the gwIMF and the IGIMF theory to calculate the changing gwIMF.

The paper is organized as follows: Section~\ref{Sec:Method} introduces the \texttt{SPS-VarIMF} code,  which calculates the luminous properties of galaxies by focusing on the varying gwIMF and the IGIMF theory. Section~\ref{sec:Results} presents the findings on how the stellar $M/L_X$ ratio, luminosity, color, and spectral energy distribution of a composite stellar population\footnote{With "composite" is meant multiple age and metallicity, in contrast to a mono-age and equal metallicity "simple" population of a star cluster for example.} evolve in both the canonical gwIMF and IGIMF framework. Finally, the paper concludes with an outlook and summary in Section~\ref{Conc}. The here-developed \texttt{SPS-VarIMF} code is made publicly available (Sec. 2.3)\footnote{https://github.com/ahzonoozi/SPS-VARIMF}.

\section{Method}\label{Sec:Method}

In this section, we describe the main ingredients necessary for SPS modeling in the framework of the varying gwIMF. The stellar evolution prescription and the stellar spectral library are described briefly considering the importance of the gwIMF on the integrated light of galaxies.

\subsection{Stellar Evolution}

Stellar evolution calculations of stars with different masses from the hydrogen burning limit ($\approx 0.1$ M$_{\odot}$) to the maximum stellar mass (150 M$_{\odot}$) give the properties of stars of any mass from the zero-age main sequence (MS) to advanced evolutionary stages. From this, it is possible to construct isochrones by specifying the location of a coeval population of stars in the  Hertzsprung-Russell (HR) diagram. 

Here we used the isochrones derived from the PARSEC \citep{Bressan2012, Marigo2017} sets of stellar evolutionary tracks, which are publicly available. The here adopted PARSEC\footnote{http://stev.oapd.inaf.it/PARSEC/} tracks  extend to stars of mass 300 $\rm M_{\odot}$ and include the pre-main sequence phase of stellar evolution.

These tracks cover all phases of stellar evolution. The isochrones are calculated for ages of $10^{5.5}-10^{10.15}$ yr with logarithmic intervals of 0.05. Isochrones are available for populations with different metallicities in the $0.0001 < Z <0.040 $ range.

The main-sequence lifetime of a 0.5 M$_{\odot}$ star is nearly 10 times longer than the Hubble time. Therefore, the low-mass stars remain on the MS without any significant evolution over the age of the universe (but see \citealt{Mansfield2023}). Although these low-mass stars contribute significantly to the total stellar mass of a galaxy, their contribution to the total luminosity is negligible. 
Stellar remnants (e.g. white dwarfs, neutron stars, or black holes) can significantly contribute to a galaxy's total stellar mass, depending on its star formation history and stellar gwIMF.

In this study, we utilize the initial-final mass relation from \cite{Spera2015} to determine the remnant masses of deceased stars. According to this approach, the initial-final mass relation is dependent on metallicity  \citep{Marigo2001, Heger2003}, which further complicates the analysis.
However, here we adopt the simple metallicity-independent prescription from \cite{Renzini1993} to assign remnant masses to dead stars: Stars with i) initial masses $m \geqslant 40 \rm M_{\odot}$ turn to black holes with a mass of $0.5 ~m$, ii) initial masses $8.5 M_{\odot}  \leqslant m < 40 M_{\odot}$ leave behind a $1.4 M_{\odot}$ neutron star after death, and finally iii) initial masses $m < 8.5 M_{\odot} $ leave a white dwarf of mass $0.077~ m + 0.48 M_{\odot}$.

\subsection{Spectral Libraries}

Using stellar spectral libraries, it is possible to convert the calculated properties of individual stars such as temperature and surface gravity to an observable spectral energy distribution. We use the semiempirical BaSeL3.1 library \citep{Lejeune1997, Lejeune1998, Westera2002}, which contains model atmosphere calculations for stars in a wide range of effective temperature and surface gravity and the metallicity range $10^{-3} \leqslant Z/Z_{\odot} \leqslant 3$, covering the full range of metallicities in the PARSEC track libraries. The library covers the whole wavelength range 91 \AA – 160 $\rm \mu m$  at resolving power $\lambda/\Delta \lambda \approx 200-500$.

\subsection{The canonical stellar IMF and the IGIMF}

The stellar IMF on the embedded cluster scale ($\approx1$ pc)  is one of the key ingredients of SPS models. Interpreting observable quantities to physical properties of a galaxy depends on the choice of the stellar IMF and how the embedded-cluster scale propagates through to the galaxy-wide scale.  For example, extracting the star formation rate of galaxies from UV or H$_{\alpha}$ emission or estimating the total stellar mass of composite stellar populations from the observed luminosity are affected substantially by the adopted stellar IMF. To investigate the importance of the stellar IMF, we compare the observational quantities of modeled galaxies being calculated with the invariant canonical gwIMF with those being calculated with the IGIMF as the galaxy-wide-IMF. 


The IGIMF theory \citep{Kroupa03} is based on the assumption that all stars form in embedded clusters \citep{Kroupa1995a, Kroupa1995b, Lada03, Kroupa05, Megeath16, Dinnbier2022} and all the clusters, whether they have been dissolved or not, form the stellar population of the galaxy. Accordingly, the gwIMF is obtained by summing the stellar IMFs of all formed embedded clusters in a given time interval $\delta t$.
The stellar IMF can be described as a multi-power-law function as formulated in \cite{Kroupa01, Kroupa02}, \cite{Yan2020, Yan2021} and \citet{Haslbauer2024}.  For the stellar IMF, we have 

\begin{eqnarray}
\xi_\star(\rm m)=\rm k_\star\left\{
      \begin{array}{ll}
\,2 \rm m^{-\alpha_1},      & 0.08 \rm M_\odot\leq m < 0.5 \rm M_\odot,  \\
\, \rm m^{-\alpha_2},      & 0.5 \rm M_\odot\leq \rm m < 1.0 \rm M_\odot,\\
\, \rm m^{-\alpha_3},      & 1.0 \rm M_\odot\leq \rm m < \rm m_{\rm max},\\
       \end{array}
        \right. \label{MF}
\end{eqnarray}
where $k_\star$ is the normalization constant.
The number of stars that form in the mass interval $m$ to $m+dm$ is $dN_\star=\xi_\star(m)dm$. The maximum mass star that can form in an embedded cluster depends on the mass in stars of the embedded cluster, $m_{\rm max}=m_{\rm max}(M_{\rm ecl}) \leqslant m_{\rm max*}$,  where $m_{\rm max*}=150 M_{\odot}$ is the adopted physical upper mass limit stars can be born with (see Eq. 6 below). The \textit{canonical stellar IMF} has $\alpha_1=1.3$, and $\alpha_2=\alpha_3=2.3$ \citep{Kroupa01}.  In general, $\alpha_1$ and $\alpha_2$ are functions of metallicity, while $\alpha_3$ is a function of both metallicity and the initial mass of the embedded cluster as follows


\begin{eqnarray}
\alpha_{1} = 1.3 + \Delta \alpha . (Z-Z_{\odot}),\ \ \ \ \ \ \ \ \ \ \ \ \ \ \ \ \ \ \ \ \ \ \ \ \ \ \ \ \ \nonumber\\ 
\alpha_{2} = 2.3 + \Delta \alpha . (Z-Z_{\odot}),\ \ \ \ \ \ \ \ \ \ \ \ \ \ \ \ \ \ \ \ \ \ \ \ \ \ \ \ \  \nonumber\\ 
\alpha_3=
\left\{
       \begin{array}{ll}
           2.3    & \  \  \ x<-0.87,\ \ \ \ \ \ \ \  \\
          -0.41x+1.94     &\  \  \ x>-0.87,\ \ \ \ \ \ \  \ 
          
 \end{array}
          \right. 
\end{eqnarray}     
where $\Delta \alpha=63$ was obtained by \cite{Yan2020} and \cite{Yan2021}. $Z$ and $Z_{\odot}= 0.0142$ \citep{Asplund2009} are the mean stellar metal-mass fraction of the system and the Sun, respectively. The parameter $x$ depends on the cloud clump density (final stars plus residual gas assuming a star formation efficiency of 1/3), $\rho_{cl}$, of the molecular cloud clump that forms the embedded star cluster, 

\begin{equation}
x=-0.14[Z]+0.99\log_{10}(\rho_{\rm cl}/10^6\rm {M_\odot pc^{-3}}),
\end{equation}
where $[Z]=\log_{10}(Z/Z_{\odot})$. The embedded-cluster cloud clump density is $\rho_{\rm cl}=3M_{\rm cl}/4\pi r_{\rm h}^3$, where $M_{\rm cl}$ is the total mass of the embedded cluster including both gas and stars. Assuming that stars form with the star formation efficiency of 33\% \citep{Lada03, Megeath16}, the stellar mass of the embedded cluster is $M_{\rm ecl} = M_{\rm cl}/3 $. For the initial half-mass radius of the cloud clump, we adopt the \cite{Marks12a} relation:

\begin{equation}
r_{\rm h}(\rm pc) = 0.1\times\left(\frac{M_{\rm ecl}}{M_{\odot}}\right)^{0.13}.\ \ \ \ \ \ \ \  \label{MKrelation}
\end{equation}
Therefore, one can compute $\alpha_3$  once the metallicity and mass of the star cluster are known as follows

\begin{equation}
x=-0.14[Z]+0.6 \log_{10}(\rm M_{\rm ecl}/10^6 \rm M_\odot)+2.82.
\end{equation}

The mass of the most massive star that forms in a cluster, $m_{\rm max}$, is a function of the embedded cluster mass, M$_{\rm ecl}$, and can be derived from the m$_{\rm max}-M_{\rm ecl}$ relation \citep{Weidner06, Yan2023}.  $m_{\mathrm{max}}$ depends on the birth cluster mass in stars,  $\rm M_{\mathrm{\rm ecl}}$, and can be obtained by solving simultaneously the following two equations
\begin{eqnarray}
    \rm M_{\mathrm{\rm ecl}} &=& \int_{0.08\, \rm M_{\odot}}^{\rm m_{\mathrm{\rm max}}} \rm m \xi_{*}(\rm m) dm \, \qquad \nonumber\\  
    1 &=& \int_{\rm m_{\mathrm{\rm max}}}^{150 \rm M_\odot} \xi_{*}(\rm m) \rm dm \, \qquad   \, . \label{eq:Mecl_mmax_normalization}
\end{eqnarray}

Note that the first part of Eq. \ref{eq:Mecl_mmax_normalization}  describes the normalization condition for the mass of the embedded cluster, $M_{\rm ecl}$, while the second part of Eq. \ref{eq:Mecl_mmax_normalization} indicates that only one star forms within the range from the maximum stellar mass ($m_{max}$) to the upper limit of stellar masses (150 $M_{\odot}$).

To calculate the gwIMF of all stars formed in the galaxy in the time interval $\delta t$, the IMFs in all embedded clusters need to be summed. The mass function of embedded clusters, the embedded cluster mass function (ECMF), is assumed to be a single power-low function with the slope of $\beta$ that depends on the SFR \citep{Weidner13b, Zhang99, Recchi09, Yan2017},

\begin{eqnarray}
\rm \xi_{\rm ecl}(M_{\rm ecl})=\left\{
   \begin{array}{ll}      
0,   &  \rm M_{\rm ecl} < M_{\rm ecl,min},  \\
\rm k_{\rm ecl} M^{-\beta} ,   & \rm M_{\rm ecl,min}\leq M_{\rm ecl} < M_{\rm ecl,max}(\rm SFR),\\
0,                     & \rm M_{\rm ecl,max}(\rm SFR)\leq M_{\rm ecl},\\ 
       \end{array}
        \right. \label{MF}
\end{eqnarray}
where

\begin{equation}
\beta=-0.106 \log_{10} .\psi(t) + 2, \label{beta}\\
\end{equation}
where $\psi(t)$ is the SFR in units of $M_\odot/\rm{yr}$ and $M_{\rm ecl,min}= 5 M_{\odot}$ is the lower stellar mass limit of the embedded clusters which is adopted based on the smallest observed star-forming stellar groups \citep{KroupaBovier03, Kirk12, Joncour18}, and $M_{\rm ecl,max}$ is the upper mass limit of embedded star clusters which is a function of the SFR.  Both  $M_{\rm ecl,max}$ and normalization constant, $k_{\rm ecl}$, can be determined by simultaneously solving the following equations
\begin{eqnarray}
\rm M_{\rm tot,\delta t}=\int_{M_{\rm ecl,min}}^{M_{\rm ecl,max}}\xi_{\rm ecl}(M)  M dM=\Psi(t)\delta t, \nonumber\\
\rm 1=\int_{M_{\rm ecl,max}} ^{10^9 M_\odot} \xi_{\rm ecl}(M) dM,\ \ \ \ \ \ \ \ \ \ \ \ \ \ \ \ \ \ \ \ \ \ \ \ \label{Mtot10}
\end{eqnarray}
where $M_{\rm tot,\delta t}$ is the total stellar mass that forms in the galaxy as a new population, through a single star formation epoch of duration $\delta t= 10$ Myr which constitutes about the typical lifetime of molecular clouds (see \citealt{Kroupa13, Schulz15} and \citealt{Yan2017} for a discussion and additional references). The first part of Equation 10 represents the total mass formed in 10 Myr, denoted as $M_{\rm tot,\delta t}$. The second part indicates that only one cluster is formed within the range of the maximum embedded cluster mass ($M_{\rm ecl,max}$) and the upper cluster mass limit ($10^9 M_{\odot}$; the results are not sensitive to this value as long as it is $>10^8 M_{\odot}$). Note that $dN_{\rm ecl}=\xi_{\rm ecl}(M_{\rm ecl})dM_{\rm ecl}$ is the number of embedded clusters in the mass interval of $M_{\rm ecl}$ and $M_{\rm ecl}+dM_{\rm ecl}$. With the stellar IMF and ECMF in hand, we can construct the stellar IMF of the whole galaxy via the IGIMF integral, by adding the IMFs of all star embedded clusters that form in the time interval $\delta t$,

\begin{eqnarray}
\rm \xi_{IGIMF}(m,\psi(t))=\,\,\,\,\,\,\,\,\,\,\,\,\,\,\,\,\,\,\,\,\,\,\,\,\,\,\,\,\,\,\,\,\,  \nonumber\\
\rm \int_{M_{ecl,min}}^{M_{ecl,max}(\psi(t))}\xi_{\star}(m\leq m_{max})\xi_{ecl} (M_{ecl},\psi(t)) dM_{ecl}.   \label{IGIMF}
\end{eqnarray}

The Fortran code \emph{GWIMF} is available to calculate the IGIMF as a function of SFR and metallicity \footnote{ https://github.com/ahzonoozi/SPS-VarIMF}. See also \citet{Yan2017, Yan2020, Yan2021} for a Python module for a similar purpose, and chemical enrichment computations and the related \textit{photGalIMF} code that in addition allows the calculation of the photometric properties of the same modelled galaxies \citep{Haslbauer2024} .

\begin{figure}
\includegraphics[width=\linewidth]{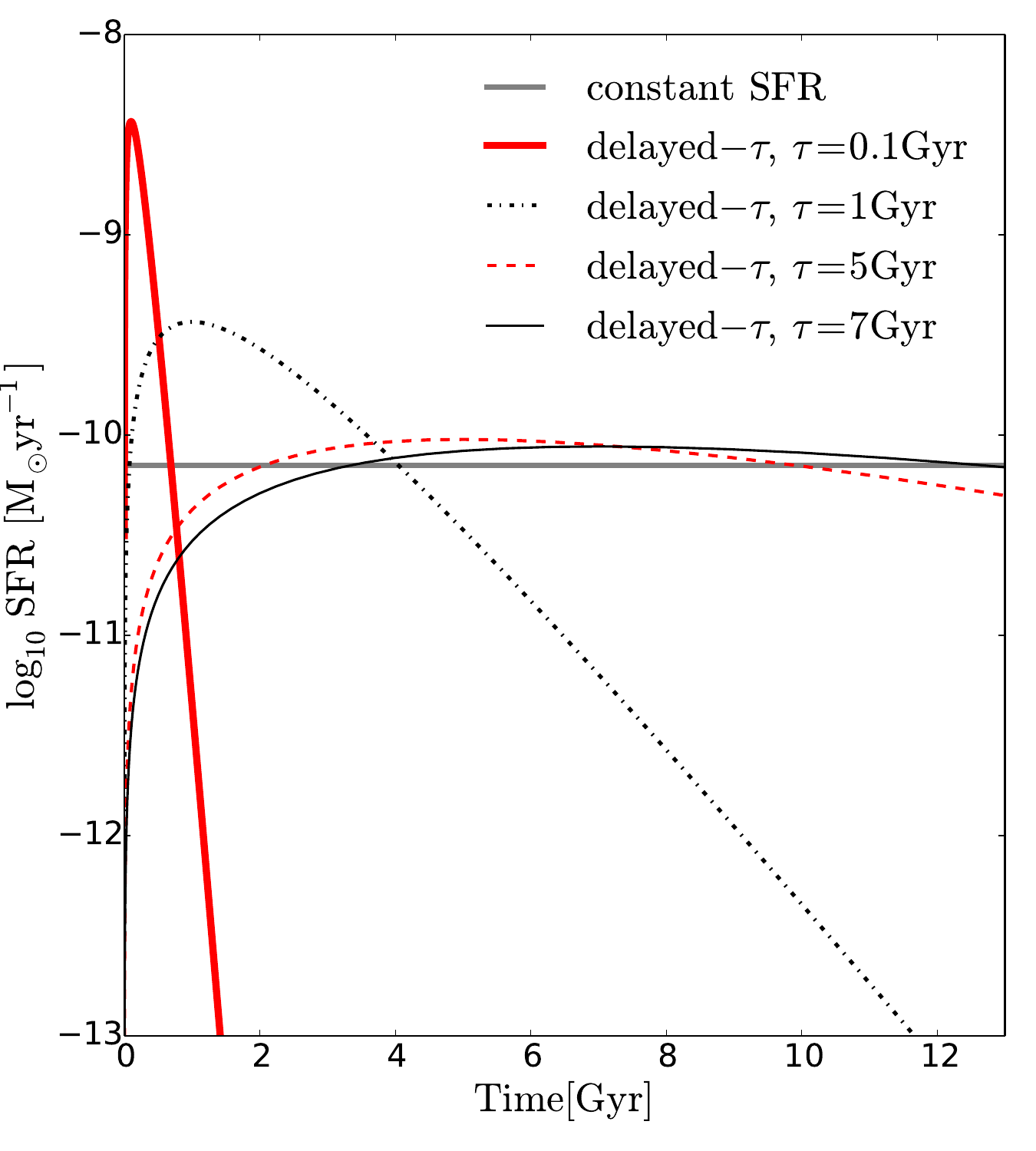}
\caption{Different SFR models used in this work (Sec. 2.4). The solid-gray line represents the constant SFR, while the different curves correspond to delayed-$\tau$ SFH models with e-folding time-scales of 0.1, 1, 5, and 7 Gyr. The area under each curve is the total stellar mass formed over 13 Gyr normalized to 1 $M_{\odot}$. The horizontal axis refers to the time since the onset of star formation.} 
\label{L_T_SSP}
\centering
\end{figure}

\subsection{Constructing the composite stellar population (CSP) of galaxies}

Combining the stellar evolution and stellar spectral libraries, and the gwIMF, we are in the position of constructing the spectrum of a single stellar population (SSP) with the metalicity of $Z$ at time $t$ after birth as follows

\begin{equation}
\rm f_{\rm SSP}(\lambda,t,Z)=\int_{m_{\rm low}}^{m_{\rm up}(t)} f_{\rm star}[\lambda,m,g_{\rm eff},T_{\rm eff},t,Z]\xi_{\rm gw}(m,t,Z)dm, \label{fssp-equation1}
\end{equation}
where $m$ is the initial (zero-age main sequence) mass of a star, $f_{\rm star}$ is its stellar spectrum that can be derived from spectral libraries once we know the metallicity $Z$, the effective temperature $T_{\rm eff}$ and the surface gravity $g_{\rm eff}$ of a star of age $t$. The IMF is denoted by  $\xi_{\rm gw}(m, t, Z)$ that can be chosen to be the invariant canonical gwIMF, $\xi_{*}(m)$, or a time-varying gwIMF according to the IGIMF($\xi_{\rm gw}=\xi_{\rm IGIMF} (m,t,Z)$). It should be noted that the IGIMF can vary with time assuming a non-constant SFR or adopting a time-dependent metallicity for the galaxy. The lower limit of integration, $\rm m_{low}=0.08 M_\odot$ is the hydrogen burning mass limit. The upper stellar mass limit, $m_{\rm up}$, is determined by stellar evolution and $t$. Both $\rm T_{\rm eff}$ and $\rm log_{10} (g_{\rm eff})$ depend on the initial mass, metallicity, and age of a star. With the spectral evolution of a coeval stellar population in hand, we can add SSPs with different ages which are weighted based on the adopted SFH to construct the time-dependent spectrum of a galaxy as follows
 
\begin{equation}
f_{\rm CSP}(\lambda,t)=\int_{t'=0}^{t'=t} \psi(t-t')f_{\rm SSP}(\lambda,t',Z) e^{-\tau_{\lambda}(t')} dt',
\end{equation}
where $\psi$ is the star formation rate, $f_{SSP}(\lambda,t',Z)$ is the spectrum of an SSP with the age $t^\prime$  and metallicity of $Z$ (eq. \ref{fssp-equation1}). The time-dependent dust attenuation is described by  $e^{-\tau_{\lambda}(t')}$ with $\tau_{\lambda}(t')$ being the effective optical depth of dust which is proportional to $\lambda^{-0.7}$ as follows

\begin{eqnarray}
\tau_{\lambda}(t')=\left\{
   \begin{array}{ll}      
\hat{\tau}_{\rm V}(\lambda/5500)^{-0.7}, ~~~\textrm{for}~~~ t'\leq 10^7  \textrm{yr},  \\
\mu_{\rm D} \hat{\tau}_{\rm V}(\lambda/5500)^{-0.7},~~\textrm{for}~~ t' > 10^7 \textrm{yr},\\ 
       \end{array}
        \right. \label{tau}
\end{eqnarray}
where $\hat{\tau}_{V}$ is the V-band optical depth in the birth clouds. The typical lifetime of giant molecular clouds is assumed to be 10 Myr. Stars younger than this age are embedded in the clouds. Once the molecular clouds are disrupted, the total dust absorption optical depth decreases instantaneously by a factor of approximately 3. The values of $\hat{\tau}_{V}=1.0$ and $\mu_D=0.33$ are used based on observational data \citep{Charlot2000, Bruzual03}.

Although the SFH of a galaxy could be very complex, we adopt the most simple and popular forms as follows:

\begin{itemize}
    \item Constant SFR which is a reasonable assumption as the time average and present-day SFR of galaxies in the local volume are nearly the same \citep{Kroupa2020, Haslbauer2023}. The analysis of 8 nearby galaxies by \citet{Yan2024b} suggests their SFR to have been very constant over the past $10^{10}$ Gyr.
    \item Exponentially declining SFR based on the closed-box model \citep{Schmidt1959},

\begin{equation}
 \psi(t)= \psi_0  e^{-t/ \tau},\\
\end{equation}
where $t$ is the age of the galaxy, $\Psi_0$ is the SFR at the beginning of the burst and  $\tau$ is the e-folding time-scale. 
    \item The delayed-$\tau$ model that allows an early phase of a rising SFR with a subsequent exponential decline (e.g., \citealt{Kroupa2020}), 
\begin{equation}
 \psi(t)= (A/\tau^2)t  e^{-t/ \tau},
\end{equation}
where $A$ is the normalization constant. All stellar mass ever formed in a galaxy, $M_{\rm tot}$, is defined as follows

\begin{equation}
\rm  M_{\rm tot}= \int_{t=0}^{t=\rm galaxy~age} \psi(t) dt. \label{Mtot}
\end{equation}

Figure \ref{L_T_SSP} compares these different star formation scenarios for a galaxy. The constant SFR and  delayed-$\tau$ SFR (ranging from $\tau=$0.1 to 7 Gyr) are shown. The total stellar mass formed over 12.5 Gyr is normalized to one solar mass.


\end{itemize}

\section{Results}\label{sec:Results}

In this section, we present the impact of the metalicity and SFH of galaxies on their present-day stellar $M/L_X$ ratios and colors with a wide range of masses in both the invariant canonical gwIMF and IGIMF contexts.

\subsection{Stellar $M/L$ ratio}

It is well known that assuming an invariant canonical stellar IMF, most of the luminosity comes from newly born high-mass stars, while most of the mass resides in low-mass stars.  However, in the context of the IGIMF, since the shape of the IMF depends on the SFH, the contribution of stellar remnants to the total galaxy mass can be significant, especially for massive galaxies with a very rapidly decreasing SFR \citep{Yan2021, Dabring2023}. On the other hand, in very low-mass galaxies with a small SFR, few if any high-mass stars form, which means that intermediate-mass stars dominate the light of the entire galaxy.

Here we calculate the stellar $M/L_X$ ratio of galaxies in different total luminosity bands, $X$. We aim to investigate how the $M/L_X$ ratio varies among the galaxies of various masses, metallicities, and SFHs.  Adopting the IGIMF theory, the galaxy-wide-IMF varies systematically with the galaxy’s star formation rate and the metallicity. This means that in the IGIMF context, not only the functional form of the SFR but also the total mass of the galaxy plays an important role in its present-day $M/L_X$ ratio.  The normalization factor of the SFR, which defines the total mass of the galaxy, is important because it affects the shape of the gwIMF and the fraction of high-mass and low-mass stars that form at a given time in a galaxy.

\begin{figure}
\includegraphics[width=8.5cm, height=8cm]{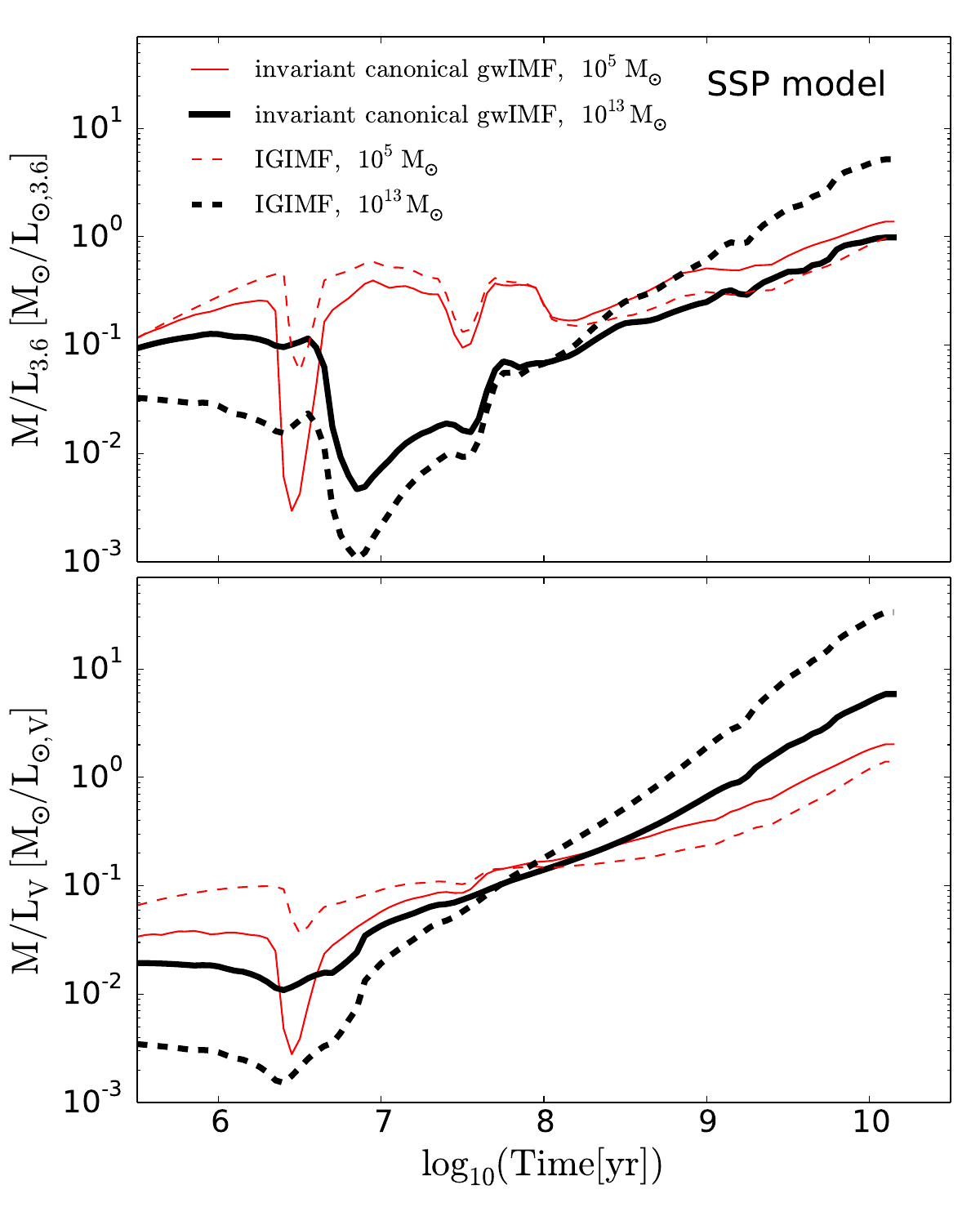}
\caption{The evolution of the $M/L_{[3.6]}$ (upper panel) and $M/L_V$ (lower panel) ratio including remnants for single stellar population (SSP) models based on the invariant canonical gwIMF and the IGIMF. Metallicity is assigned to galaxies with different masses following the mass-metallicity relation of \protect\cite{Gallazzi2005} such that $Z=0.03$ for the $M_{\rm tot} = 10^{13} M_{\odot}$ model and $Z= 0.0002$ for the $M_{\rm tot}=10^5 M_{\odot}$ model. The horizontal axis refers to the time since the stellar population was born.} 
\label{ML_T_SSP}
\centering
\end{figure}

\begin{figure}
\includegraphics[width=8.5cm, height=10cm]{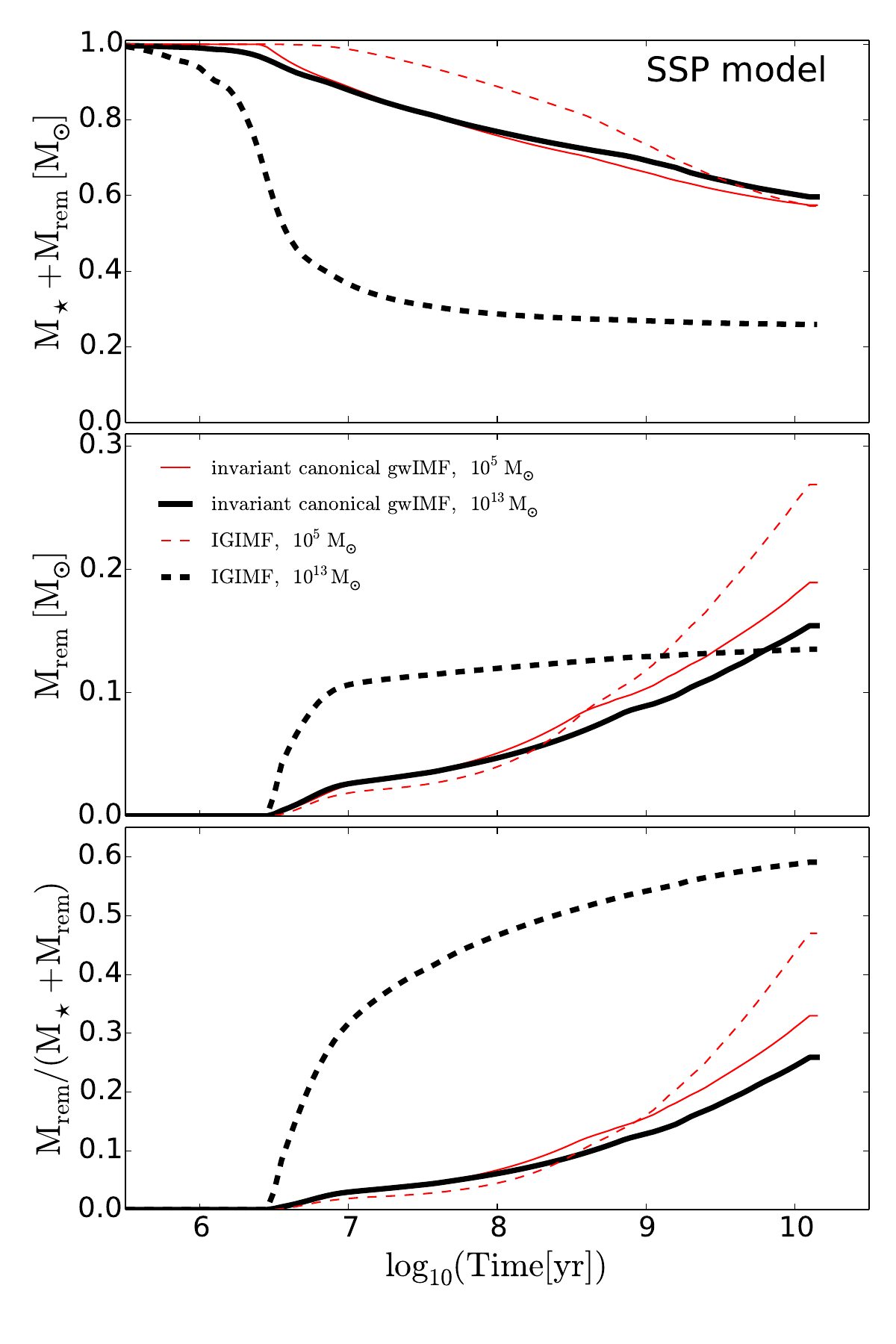}
\caption{The evolution of the total stellar mass including the live stars plus remnants (upper panel) and remnants only (middle panel) and the remnant fraction (lower panel) for single stellar population (SSP) models based on the invariant canonical gwIMF and the IGIMF. Metallicity is assigned to galaxies with different masses following the mass-metallicity relation of \protect\cite{Gallazzi2005} such that $Z=0.03$ for the $M_{\rm tot} = 10^{13} M_{\odot}$ model and $Z= 0.0002$ for the $M_{\rm tot}=10^5 M_{\odot}$ model. The horizontal axes refer to the time since the stellar population is born.  The total mass of the galaxies is normalized to $1 M_{\odot}$. To obtain the actual mass of the galaxies at each point, multiply the values by the corresponding mass $M_{\rm tot}$ indicated in the figure.}
\label{Mass_T_SSP}
\centering
\end{figure}

\subsubsection{Simple stellar population (SSP)}
The time evolution of the $M/L_X$ ratio in the $X=[3.6]$-band (upper panel) and $X=V$-band (lower panel) for an SSP constructed as a single-age population born at $t=0$ based on the invariant canonical gwIMF and the IGIMF are compared in Figure \ref{ML_T_SSP}. Galaxies with two extreme masses, $M_{\rm tot}= 10^5 M_\odot$ and $M_{\rm tot}= 10^{13} M_\odot$ are adopted and metallicity is assigned to each population following the mass-metallicity relation \citep{Gallazzi2005}.  Note that in general $M_{\rm tot}$ is the total mass of all stars formed in the galaxy (i.e. the integral of the SFH). Therefore, in the context of the invariant canonical gwIMF, the resulting difference in the $M/L_X$ ratio of models with different galaxy masses is due to the difference in metallicity. The stars' metallicity significantly impacts their color and brightness, especially in the optical and ultraviolet bands. Metal-poor stars tend to have a bluer color and emit more UV radiation compared to metal-rich stars. The effect of metallicity becomes more important when studying stellar populations of galaxies. Galaxies of different ages have different metallicities due to self-enrichment. Different regions of a galaxy can have different metallicities, which can affect the observed color and luminosity of those regions.

As shown in Figure \ref{ML_T_SSP}, metallicity plays a more important role in the $V$-band luminosity (bottom panel) compared to the $[3.6]$-band (top panel). Metal-poor galaxies emit a larger fraction of their bolometric luminosity in the V-band than metal-rich galaxies. For the gwIMF the being invariant canonical IMF, this would result in a metal-rich galaxy ($M_{\rm tot} = 10^{13} M_{\odot}$ and  $Z=0.03$)  having a present-day value of $M/L_V=6 M_\odot/L_\odot$ at $t=12.5$ Gyr which is larger than for a low mass galaxy ($M_{\rm tot}=10^5 M_{\odot}$  and $Z= 0.0002$) by a factor of 3, while the $M/L_{3.6}$ ratio does not change much for different metallicities. Note that the here quoted $M/L_X$ ratios include stellar remnants.

\begin{figure}
\includegraphics[width=8.5cm, height=8cm]{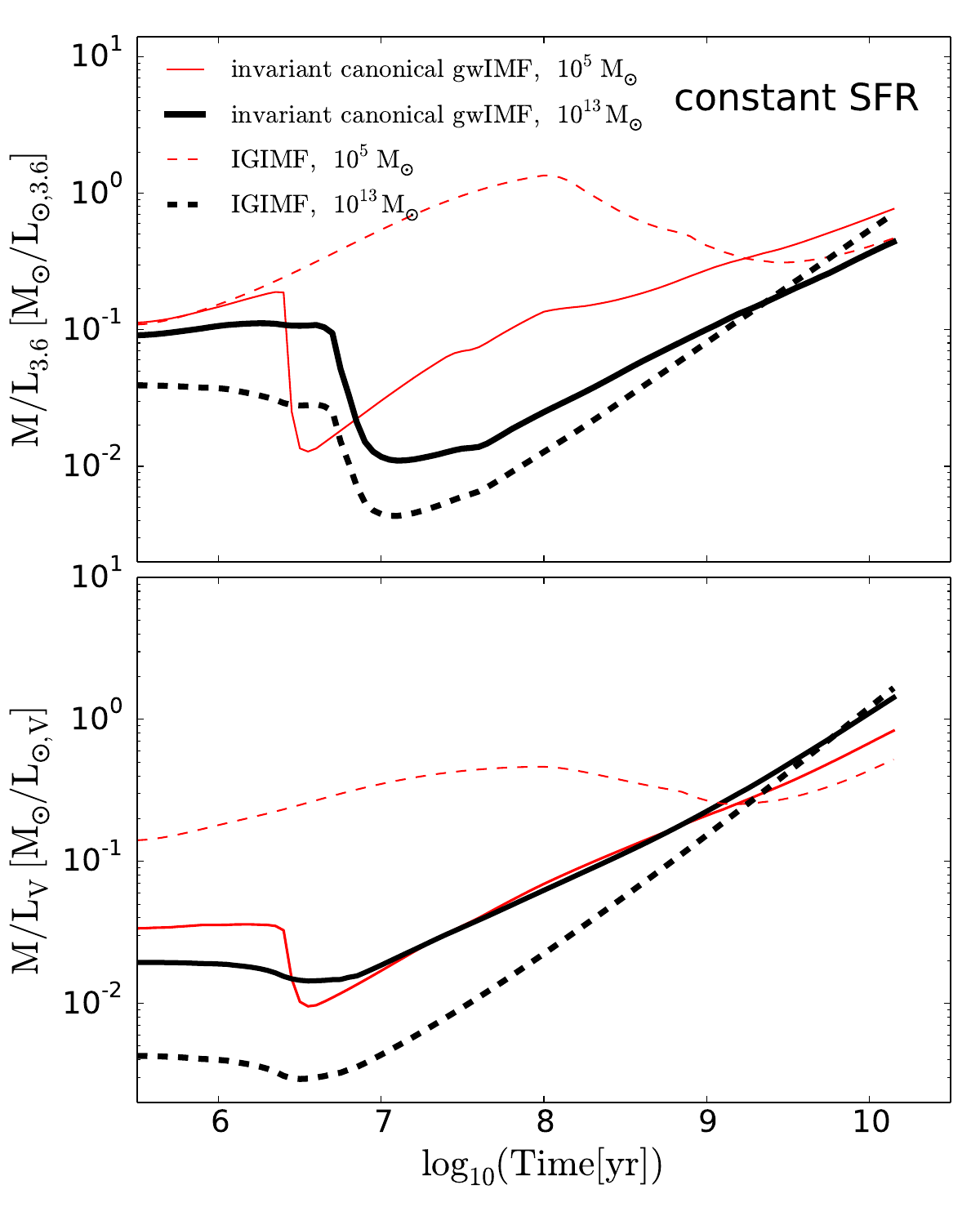}
\caption{Same as Figure \ref{ML_T_SSP} but for a CSP with a constant SFR for 12.5 Gyr. The models have SFR=$10^{-5} M_{\odot}/\rm yr$ ($M_{\rm tot}=10^5 M_\odot$ at $t=12.5$ Gyr) and SFR=$10^{3} M_{\odot}/\rm yr$ ($M_{\rm tot}=10^{13} M_\odot$ at $t=12.5$ Gyr) and are computed to their present-day mass for 12.5 Gyr.  }
\label{ML_T_const}
\centering
\end{figure}

\begin{figure}
\includegraphics[width=8.5cm, height=10cm]{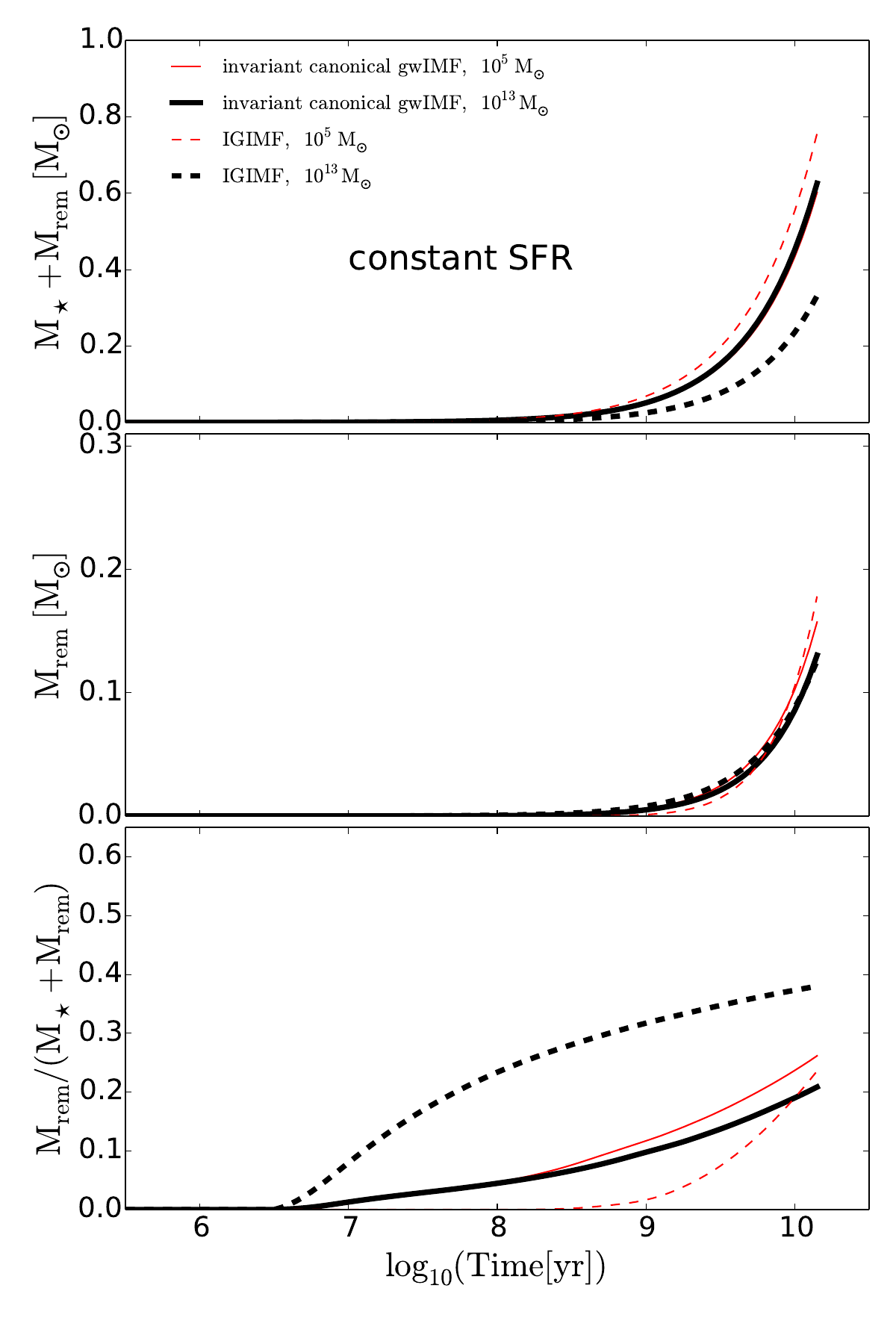}
\caption{Same as Figure \ref{Mass_T_SSP} but for a CSP with a constant SFR. The models have SFR=$10^{-5} M_{\odot}/\rm yr$ ($M_{\rm tot}=10^5 M_\odot$) and SFR=$10^{3} M_{\odot}/\rm yr$ ($M_{\rm tot}=10^{13} M_\odot$) and are computed to their present-day mass for 12.5 Gyr.}
\label{Mass_T_CONST}
\centering
\end{figure}

The difference between the $M/L_X$ ratios of massive galaxies and low-mass ones increases in the IGIMF context compared to the invariant canonical gwIMF.  As the total galaxy mass in stars changes from $M_{\rm tot}=10^5 M_\odot$ to $10^{13} M_\odot$, the present-day ($t=12.5$ Gyr) $M/L_X$ ratio varies from 1.0(1.4) to 6(40) in the $\rm [3.6](\rm V$) band. This large difference is not only due to the different assigned metallicity but also because the star formation rate plays the most important role in the evolution of the $M/L_X$ ratio and its present-day value. The SFR is larger in massive galaxies, which makes them have more top-heavy gwIMFs, according to the IGIMF theory, meaning that more massive stars have formed in these galaxies and ultimately a larger fraction of their present-day mass is in stellar remnants. Since the contribution of the stellar remnants to the luminosity is negligible, the $M/L_X$ ratio increases more rapidly for massive galaxies with the age of the stellar population such that the present-day $M/L_X$ ratios of high-mass galaxies are significantly larger than when an invariant canonical gwIMF is assumed. 

For the low-mass galaxies with small SFRs ($<1M_\odot/\rm yr$), the IGIMF becomes top-light regardless of the metallicity, and due to the relatively large number of faint low-mass stars, the initial value of the $M/L_X$ ratio is larger in the IGIMF context. However, after their long-term evolution, due to the lack of remnant stars, their present-day $M/L_X$ ratios will be smaller than those of massive galaxies. Taking into account the spread of the mass-metallicity relation, using a higher metallicity for low-mass galaxies results in a bottom-heavy gwIMF, without significantly affecting the gwIMF slope for massive stars.  Fig. \ref{Mass_T_SSP} shows the time evolution of stellar mass including live stars plus remnants ($M_* + M_{\rm rem}$, upper panel), only remnants (middle panel) and the fraction of mass in remnants (lower panel). This leads to the production of more faint low-mass stars, ultimately impacting the initial value of the $M/L_X$ ratio. The initial total mass in stars, $M_{\rm tot}$, is normalized to one solar mass. In the context of the gwIMF being the invariant canonical IMF, the metal-poor galaxies have a larger fraction of mass in remnants because metal-poor stars retained more massive remnants (see Sec. 2.1 for more details). In the IGIMF context, for massive galaxies, the total mass of the remnants increases at around a few Myr and then grows slowly. This indicates that, based on the IGIMF theory, the massive galaxies are dominated by BHs and NSs formed in the first few Myr. 
 The fraction of mass in remnants, $f_{\rm rem}=M_{\rm remn}/(M_*+M_{\rm rem})$ is shown in Fig. \ref{Mass_T_SSP}.

\subsubsection{Composite stellar population (CSP)}

Star formation is a complex process that can proceed in different ways in different types of galaxies. The rare giant elliptical galaxies formed in a single intense burst of star formation about $10^{10}$ years ago and were quenched after that \citep{Cowie1996, Thomas2005, Recchi09, McDermid2015, Yan2021, Eappen2022}, while the common spiral galaxies evolve slowly and still have ongoing star formation \citep{Kroupa2020}. 
Therefore, it is useful to adopt different SFHs to investigate the $M/L_X$ ratios of various types of galaxies.

Assuming that the SFR is constant, we calculate how the $M/L_X$ ratios of different CSPs vary with time if their stars form following the invariant canonical gwIMF or the IGIMF. As illustrated in Figure \ref{ML_T_const}, for the invariant canonical gwIMF, the current value of $M/L_{[3.6]}$ is 0.7 (for $M_{\rm tot}=10^5 M_\odot$) to approximately 0.4 (for $M_{\rm tot}=10^{13} M_\odot$) which is due to the difference in metallicity. Metal-poor stars are predicted to lose less mass than metal-rich stars because of their weaker stellar winds. This results in metal-poor stars leaving massive remnants at the final stage of their evolution. Consequently, metal-poor galaxies have larger $M/L_{[3.6]}$ values. According to the IGIMF theory, the present-day $M/L_{[3.6]}$  ratios are from 0.5 (for $M_{\rm tot}=10^5 M_\odot$) to $0.7$ (for $M_{\rm tot}=10^{13} M_\odot$).

As can be seen in Figure \ref{ML_T_const}, massive stars evolve into supergiants and convert to black holes and neutron stars through supernova explosions at around a few Myr. This leads to an increase in the total luminosity (i.e., a decrease in the $M/L_{[3.6]}$ ratio) of the galaxy, and then a decrease at the time after converting the luminous massive stars into dark remnants (i.e., the $M/L$ ratio in both bands increases). However, the evolution of the low-mass galaxy with $M_{\rm tot}=10^5 M_{\odot}$ in the IGIMF theory is different, as there is no drop of the $M/L_V$ ratio around the age of a few Myr. This is because of the low value of the SFR which is  $SFR=M_{\rm tot}/12.5 \rm ~Gyr \approx10^{-5} M_{\odot}/yr$ for a galaxy with  $M_{\rm tot}= 10^5 M_{\odot}$. In the IGIMF theory, stellar populations that form with such a small SFR would not contain any stars with  $m>4 M_{\odot}$ (see \citealt{Yan2017}). Therefore, such a low-mass galaxy does not experience the evolution of massive stars around a few Myr. Instead, they show such behavior at later times, at around 100 Myr. This is when stars with a mass of about $4 M_{\odot}$  evolve from the main sequence to advanced stages of stellar evolution. Thus the luminosity rises through RGB and AGB phases and therefore the $M/L_{[3.6]}$ ratio drops. After these stars convert to remnants, the $M/L_{[3.6]}$ ratio increases again due to the cumulative mass of remnant stars. Figure \ref{Mass_T_CONST} illustrates the time evolution of stellar mass, including both live stars and remnants ($M_* + M_{\rm rem}$, upper panel) as well as remnants only (middle panel) and the fraction of mass in remnants (lower panel) for a constant SFH.

\begin{figure}
\includegraphics[width=8.5cm, height=8cm]{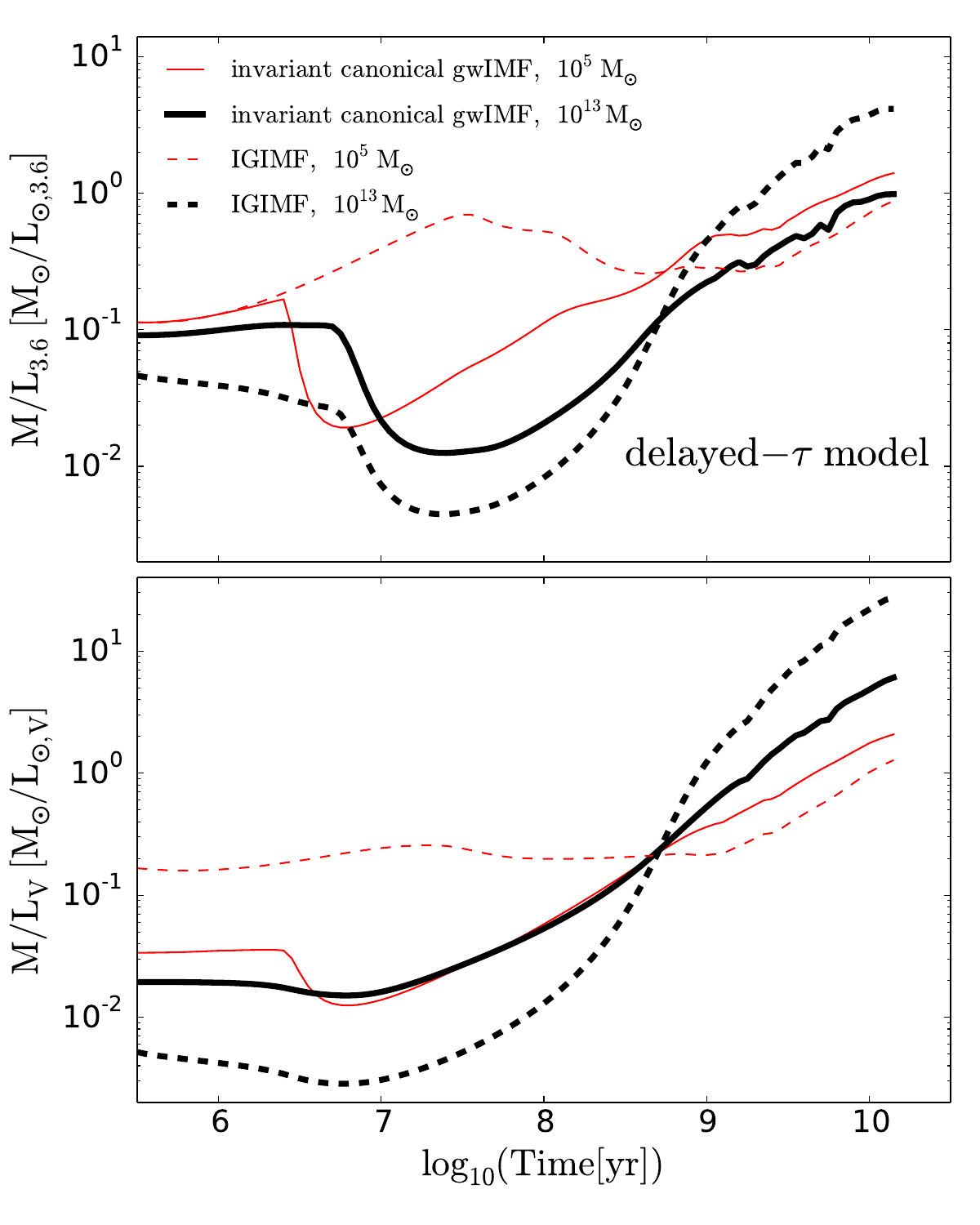}
\caption{Same as Figure \ref{ML_T_SSP} but for a CSP with a delayed-$\tau$ SFH.}
\label{ML_T_Delay-1}
\centering
\end{figure}

\begin{figure}
\includegraphics[width=8.5cm, height=10cm]{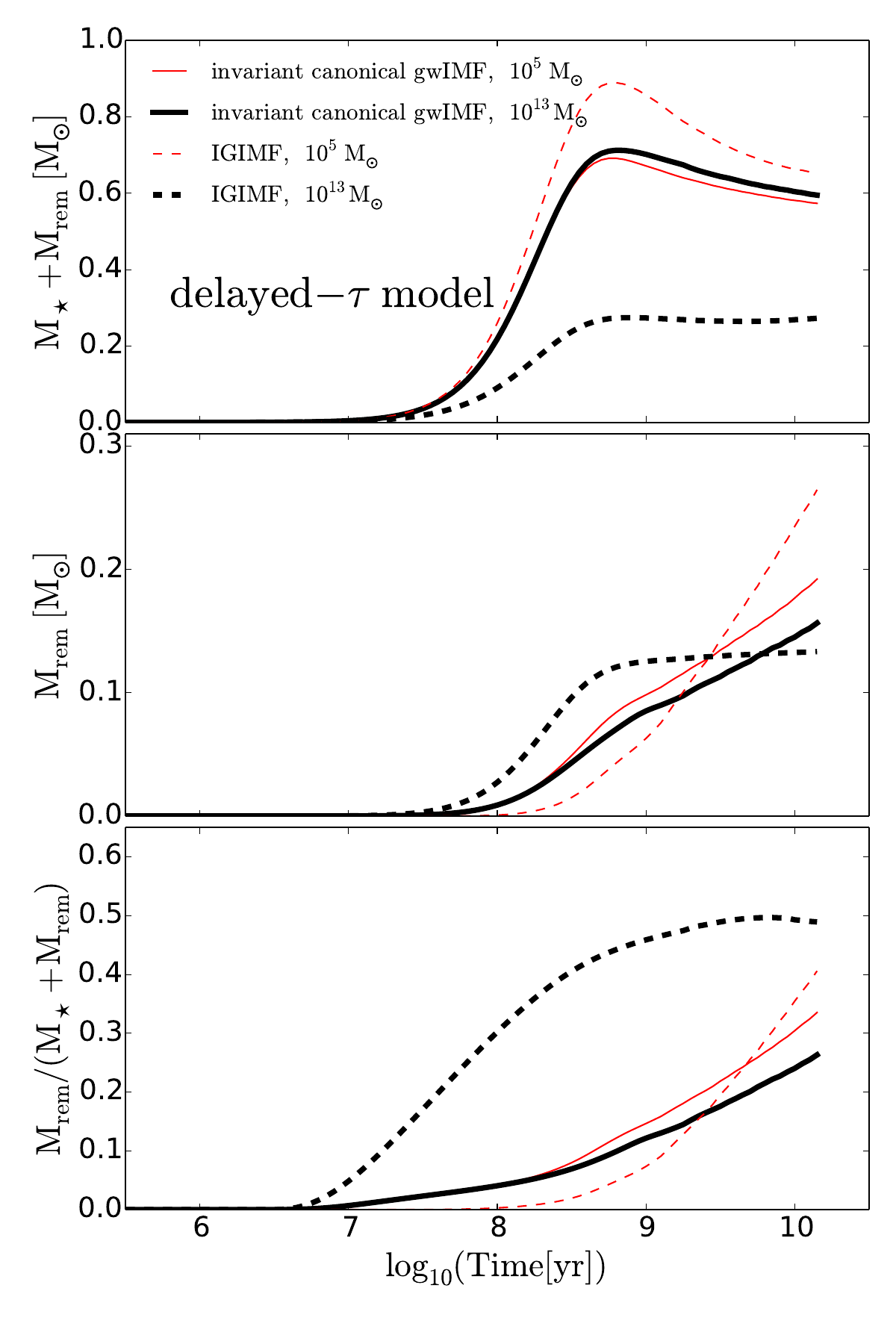}
\caption{Same as Figure \ref{Mass_T_SSP} but for a CSP with a delayed-$\tau$ SFH. }
\label{Mass_T_Delay}
\centering
\end{figure}

For comparison with the above, a set of models with the delayed-$\tau$ SFH (Sec. 2.4) are also calculated. To explore the effect of quenching, we adopt a relatively short value of $\tau=0.1$ Gyr, which leads to a higher and earlier SFR and consequently a larger mass in remnants. These are models for elliptical galaxies. Especially for massive galaxies within the IGIMF theory, the fraction of remnant stars would be significant as early high SFR leads the early gwIMF to be populated by more massive stars (i.e., gwIMF is top-heavy). So the large fraction of remnant stars and the lack of luminous massive stars due to the present-day top-light gwIMF (due to the decreasing SFR at late times) imply these galaxies have very large $M/L$ ratios in the V-band and at 3.6 $\rm \mu m$ (Figure \ref{ML_T_Delay-1}). The present-day ($t=12.5$ Gyr) $M/L_X$ ratios of galaxies with different masses and SFHs are tabulated for the canonical gwIMF and the IGIMF in Table 1. Moreover, for comparison we plot the time evolution of stellar mass, including both live stars and remnants ($M_* + M_{\rm rem}$, upper panel) as well as remnants only (middle panel) and fraction of mass in remnants (lower panel) for the delayed-$\tau$ SFH  (Fig. \ref{Mass_T_Delay}).

\begin{table*}
	\centering
	\begin{tabular}{cccccc}

		\hline  
$M_{\rm tot}[M_{\odot}]$ & SFH & IMF & $[(M_*+M_{\rm rem})/M_{\rm tot}](12.5~ \rm Gyr)$  & $M/L_{[3.6]}$  & $M/L_V$  \\        
		\hline 

        $10^5$     & SSP            & canonicIMF &0.57& 1.4 & 2.0  \\
        $10^5$     & SSP            & IGIMF      &0.56& 1.0 & 1.4  \\
        $10^5$     & delayed$-\tau$ & canonicIMF &0.57& 1.4 & 2.0  \\
        $10^5$     &  delayed$-\tau$& IGIMF      &0.66& 1.0 & 1.5  \\
        $10^5$     & constant       & canonicIMF &0.54& 0.9 & 0.9  \\
        $10^5$     & constant       & IGIMF      &0.68& 0.5 & 0.6 \\
        $10^{13}$  & SSP            & canonicIMF &0.59& 1.0 & 6   \\
        $10^{13}$  & SSP            & IGIMF      &0.23& 6.0 & 40.0  \\
        $10^{13}$  &delayed$-\tau$  & canonicIMF &0.59& 1.0 & 7.0  \\ 
        $10^{13}$  & delayed$-\tau$ & IGIMF      &0.27& 4.0 & 30.0  \\		
        $10^{13}$  & constant       & canonicIMF &0.56& 0.5 & 1.5  \\
        $10^{13}$  & constant       & IGIMF      &0.30& 0.7 & 1.5  \\
        
       		\hline
	\end{tabular}
	\caption{  The present-day ($t=12.5$ Gyr) $M/L_{\rm V}$ and $M/L_{[3.6]}$ ratios of galaxies with different masses and SFHs for the canonical gwIMF and the IGIMF. The delayed-$\tau$ models have $\tau=0.1$ Gyr. The $M_{\rm tot}$ is the time integral over the SFR (eq. \ref{Mtot}) and comprises the mass of all stars formed while $[M_*+M_{\rm rem}](12.5~ \rm Gyr)$ is the mass in living stars plus remnants at time $t=12.5~ \rm Gyr$. }
	\label{tab1}
\end{table*}

\subsection{Luminosity and color}

\begin{figure}
\includegraphics[width=8.5cm, height=10cm]{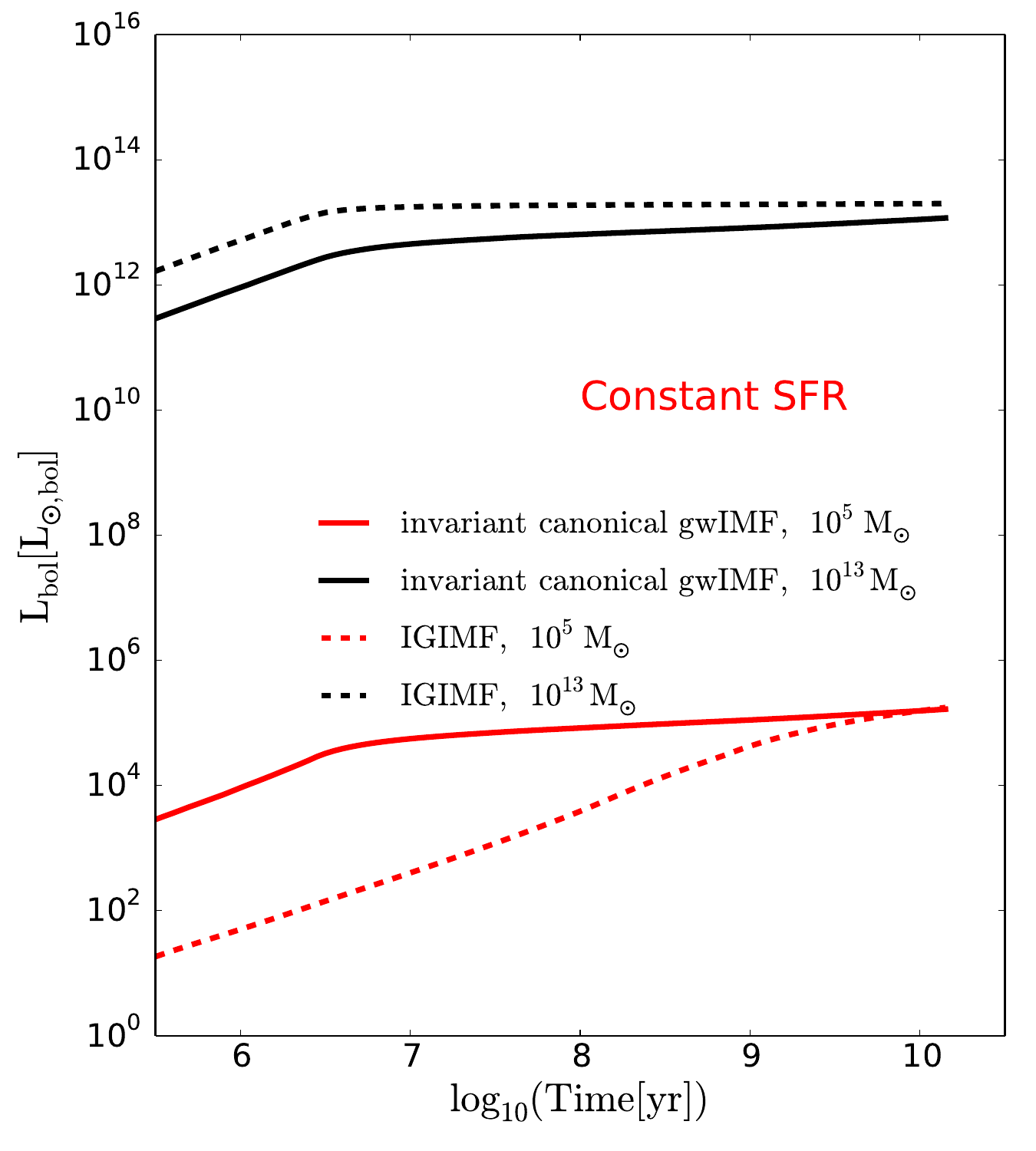}
\caption{The evolution of the total bolometric luminosity of star-forming galaxies with different masses based on the invariant canonical gwIMF and the IGIMF assuming a constant SFR. These are models of the late-type or disk galaxies.} 
\label{Bol1}
\centering
\end{figure}

\begin{figure}
\includegraphics[width=8.5cm, height=10cm]{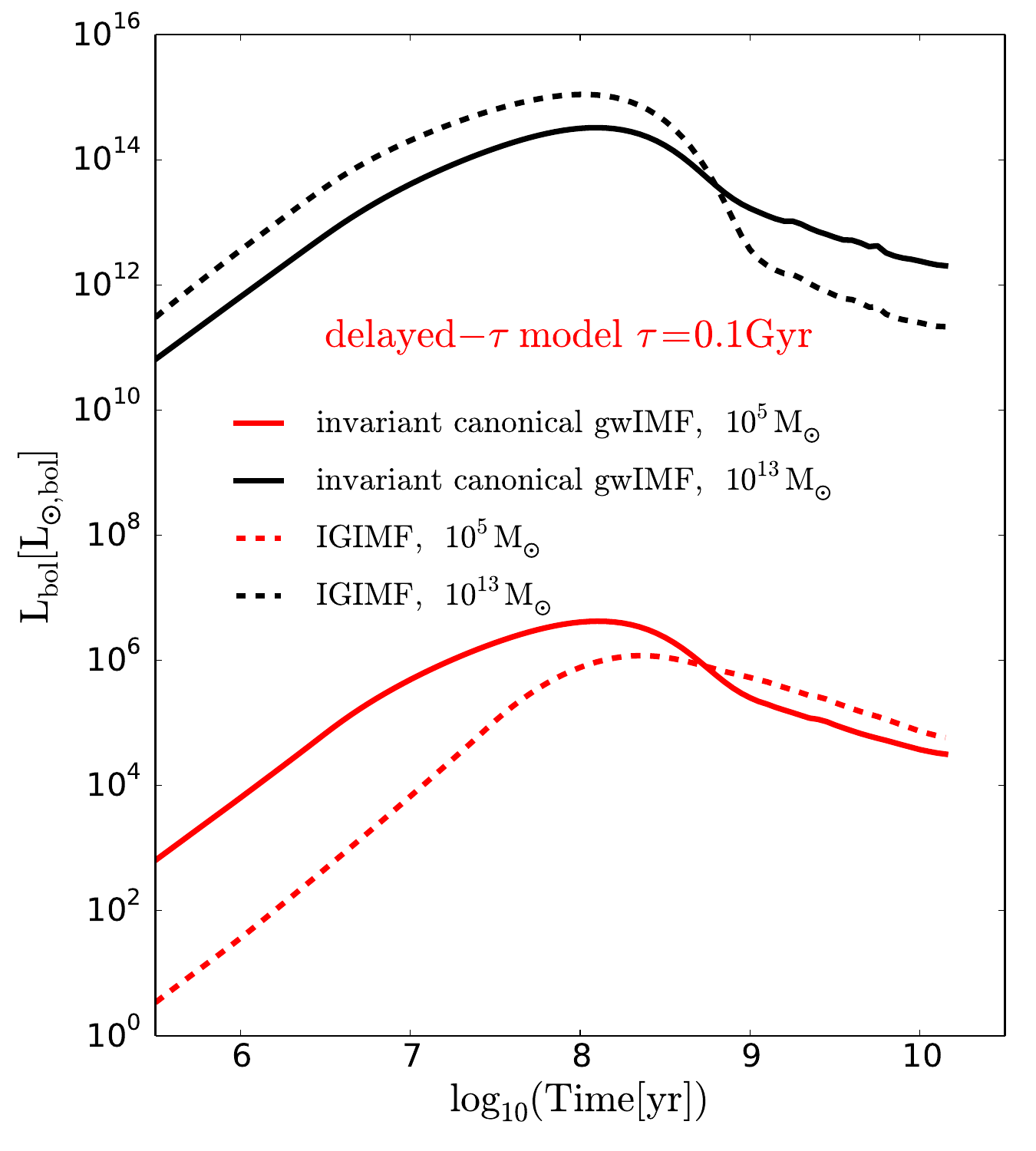}
\caption{Same as Fig. \ref{Bol1} but for early-type galaxies with a delayed-$\tau$ SFH. Note that an elliptical galaxy that has a bolometric luminosity today of $10^{12} L_{\odot, \rm bol}$ would have had $L_{\rm bol}=10^{16} L_{\odot, \rm bol}$ at its peak according to the IGIMF theory. }  
\label{Bol2}
\centering
\end{figure}

A comparison between the observed and predicted colors may help to reveal the SFH, metallicity, and gwIMF of galaxies. Different colors are sensitive to varying periods of star formation. The UV luminosities mostly come from O and B stars and represent the newly born stars over the last 100 Myr, indicating the present-day SFR \citep{Pflamm08, Schombert2019, Schombert2020}.  The bolometric luminosities of star-forming and elliptical galaxies are shown in Figures \ref{Bol1} and \ref{Bol2}, respectively. As shown, for the same $M_{\rm tot}$, massive elliptical galaxies have lower present-day bolometric luminosities in the IGIMF theory, while star-forming galaxies exhibit larger luminosities.  For completeness, the results of calculations for two other bands and colors (K- and I-band vs. V-K and V-I colors) are repeated, and the results
are shown in Figures \ref{A1} and \ref{A2}.

\begin{figure}
\includegraphics[width=8.5cm, height=8cm]{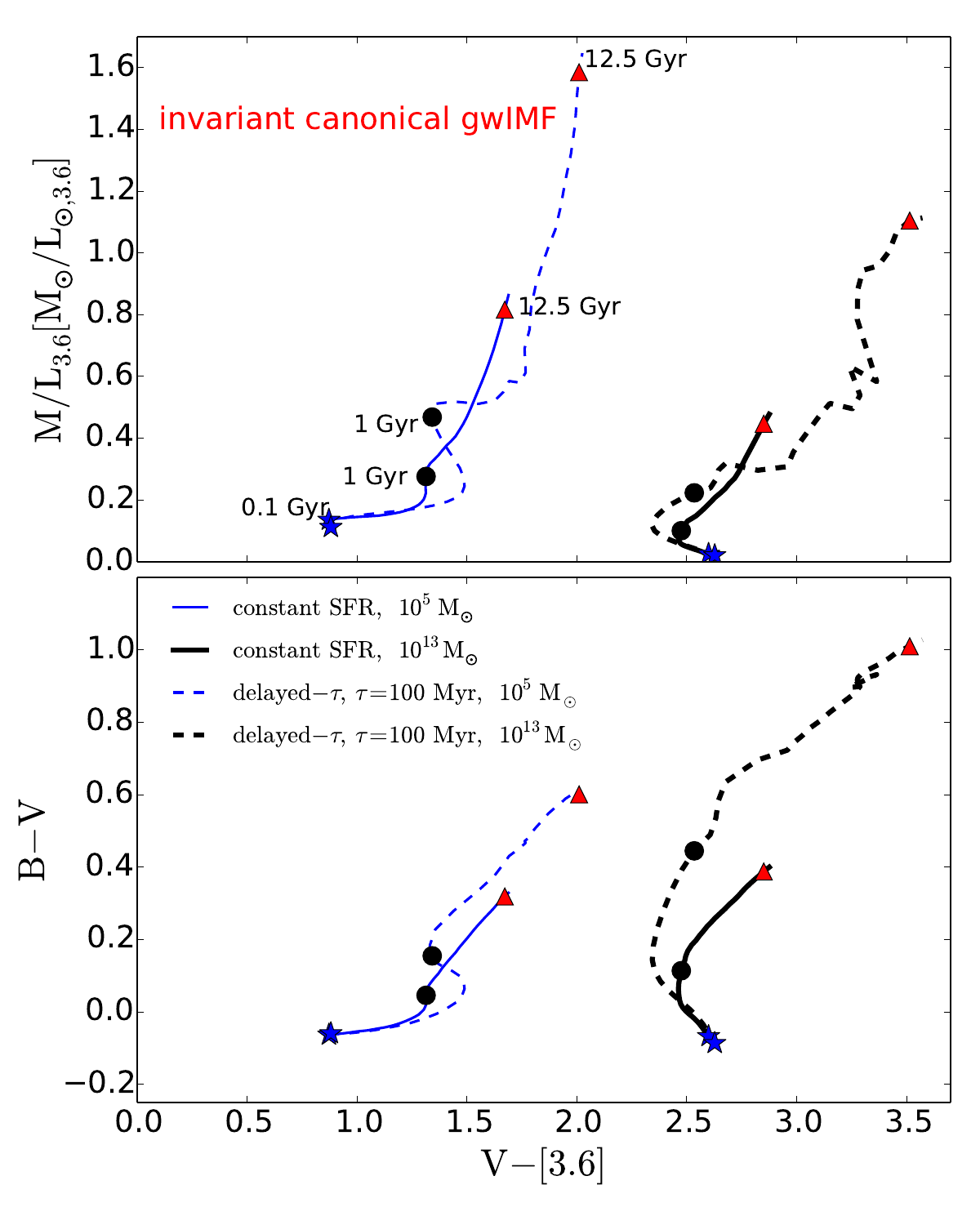}
\caption{The bottom panel displays the average optical to near-IR colors for galaxies with different masses and SFHs. Two SFR models are shown:  solid lines for constant SFR, dashed lines for delayed-$\tau$ SFR, red lines for low-mass (metal-poor) galaxies, and black lines for massive (metal-rich) galaxies.  The top panel displays the effect of those same models on the deduced $M/L_{[3.6]}$ ratios including remnants. All models are constructed based on the invariant canonical gwIMF and evolve from the lower left to the upper right. The symbols along the curves represent different ages, ranging from 0.1 to 12.5 Gyr.}
\label{ML_T_Delay}
\centering
\end{figure}

Different phases of stellar evolution, such as the RGB and AGB phases, influence the V-K color. Since the gwIMF determines the relative weights of different parts of the isochrones,  assuming different IMFs results in different colors for a galaxy which in turn will be interpreted as different physical properties such as age and metallicity. The upper panel of Figure \ref{ML_T_Delay} shows  $M/L_{[3.6]}$ vs. the V-$[3.6]$ color for galaxies with two different masses and accordingly two different metallicities, which are shown by red ($Z=0.0002$) and black ($Z=0.03$) lines. All models are constructed based on the invariant canonical gwIMF. For completeness, the results of calculations for two other bands and colors (K- and I-band vs. V-K and V-I colors) are repeated and the results are shown in Figures \ref{A3} and \ref{A4}.

We use a constant SFR and a delayed$-\tau$ SFH with $\tau=0.1 \rm Gyr$ to find the evolutionary tracks in the 2D $M/L_X$-color space. As illustrated in Figure \ref{ML_T_Delay} (upper panel) models with declining SFRs can reach higher $M/L_{[3.6]}$ ratios of 1.4 (1) compared to the constant SFR models that reach 0.7  (0.4) for metal-poor (for metal-rich)  galaxies in the declining models due to the larger masses of remnants. In addition to the color, the effect of metallicity is also important in the $M/L_{[3.6]}$ ratio as metal-poor galaxies can reach larger values of the $M/L_{[3.6]}$ ratio at the same age owing to the more massive remnants.

It is worth noting that remnant masses of metal-poor stars are larger than metal-rich ones, that intensify the dependency of the $M/L_X$ ratio on the metallicity, especially for the high-mass galaxies within the IGIMF context. 

With the increasing age of the galaxy, more stars move to the giant branch and the galaxy's color becomes redder. Also, with increasing metallicity, the opacity increases in stellar atmospheres resulting in a lower effective star temperature and thus redder colors for galaxies. Therefore, based on the mass-metallicity relation, we expect massive spiral galaxies with a larger amount of past SFR to be redder. Our models show that very old metal-rich galaxies with a fast declining SFR can have a V-$[3.6]$ color of 3.5 and a $B$-$V$ color of 1.  Also, the lack of massive luminous blue stars in declining SF models makes these galaxies redder than those with continuous constant SFR.

It should be noted that the early intense and rapid star formation (i.e., delayed$-\tau$ models with a small value of $\tau=0.1$ Gyr) enriches the gas and produces a metal-rich present-day population. But in a galaxy with a constant SFR, the chemical enrichment proceeds slowly which leads the galaxy to be dominated by metal-poor stars. For simplicity,  here we do not consider the time evolution of metallicity or its dependency on the SFH. For this, a chemical enrichment code would be needed (e.g., \citealt{Yan2021, Gjergo2023, Haslbauer2024}).

The lower panel of Figure \ref{ML_T_Delay} demonstrates the two-color diagram of B-V versus V-$[3.6]$. The optical color, B-V, is sensitive to newly born  OB stars and therefore presents the recent star formation of the galaxy. However, the V-$[3.6]$ color is sensitive to old populations and the fraction of stars evolving to the RGB and AGB phases. The impact of metallicity is stronger for the V-$[3.6]$ color which is evident from the significant amount of horizontal displacement in the V-${[3.6]}$ color (lower panel of Figure \ref{ML_T_Delay}) between metal-poor (blue lines) and metal-rich models (black lines).
The V-$[3.6]$ color lies in the range of [2.4-3.5] for massive (metal-rich) galaxies with different ages and SFHs in agreement with the observed spread in colors of massive disk galaxies (see figure 2 of \citealt{Schombert2019}). These galaxies have similar metallicities according to the flattening of the mass-metallicity relation at the high mass end.

\begin{figure}
\includegraphics[width=8.5cm, height=8cm]{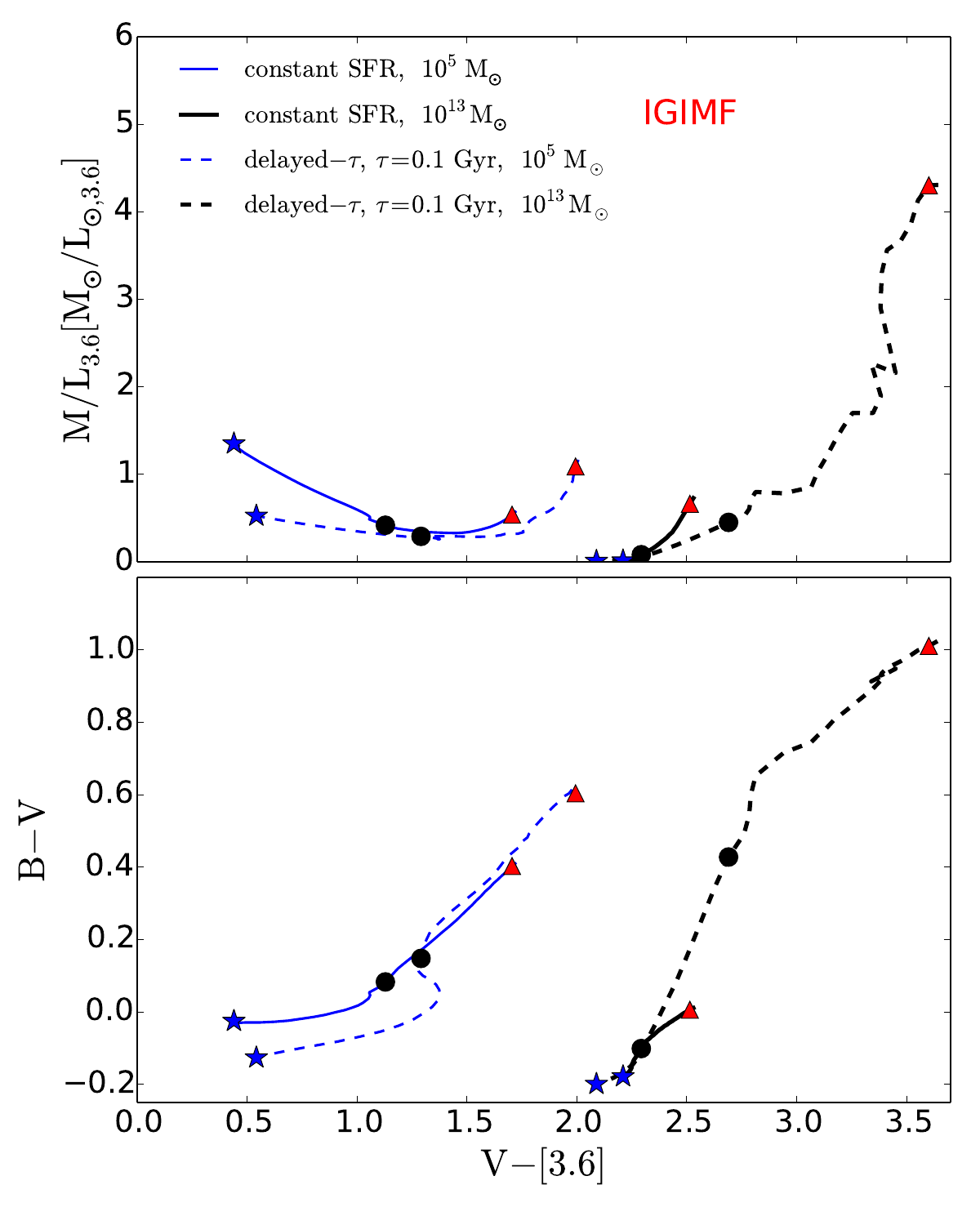}
\caption{Same as Figure \ref{ML_T_Delay} but all models are constructed based on the IGIMF theory. }
\label{ML_T_Delay-IGIMF}
\centering
\end{figure}


In Figure \ref{ML_T_Delay-IGIMF}, the calculations are repeated for models with the IGIMF to compare with the canonical gwIMF.  As can be seen, the predicted $M/L_{[3.6]}$ ratio increases drastically by a factor of about five for massive galaxies with rapidly declining SFR in comparison to the invariant canonical gwIMF case.

The maximum value of B-V that a massive galaxy (i.e., a metal-rich galaxy with $V-[3.6]=2.5$) with constant SFR can reach over the Hubble time is about 0.0  in the IGIMF  context (solid black line in the lower panel of Figure \ref{ML_T_Delay-IGIMF}), which is lower than what the same galaxy can have ($B-V=$ 0.4) based on the invariant canonical gwIMF (solid black line in the lower panel of Figure \ref{ML_T_Delay}). A decrease in SFR with time causes galaxies to appear redder within the invariant canonical gwIMF and the IGIMF contexts.

Comparing the bottom panels of Figures \ref{ML_T_Delay} and \ref{ML_T_Delay-IGIMF}, it can be seen that for massive galaxies with constant or slowly declining SFRs (i.e., solid black lines) the choice of the gwIMF has a clear effect on the colors. Therefore studying the broadband luminosities and different color bands can be used to constrain the mass and gwIMF of late-type galaxies with constant (or slowly declining) SFRs.

\subsection{Revealing the underlying gwIMF in galaxies}
In this section, we investigate the possibility of identifying the underlying gwIMF of massive galaxies through the SPS technique. In other words, the idea is that by comparing the brightness and colors of observed galaxies with the results of simulated stellar populations with different assumptions about the gwIMF, one can determine how stars are distributed within a galaxy in terms of their masses. Since luminosity is often the only observable parameter of galaxies, we compare two galaxies with the same present-day luminosities.

\begin{figure}
\includegraphics[width=9cm, height=10cm]{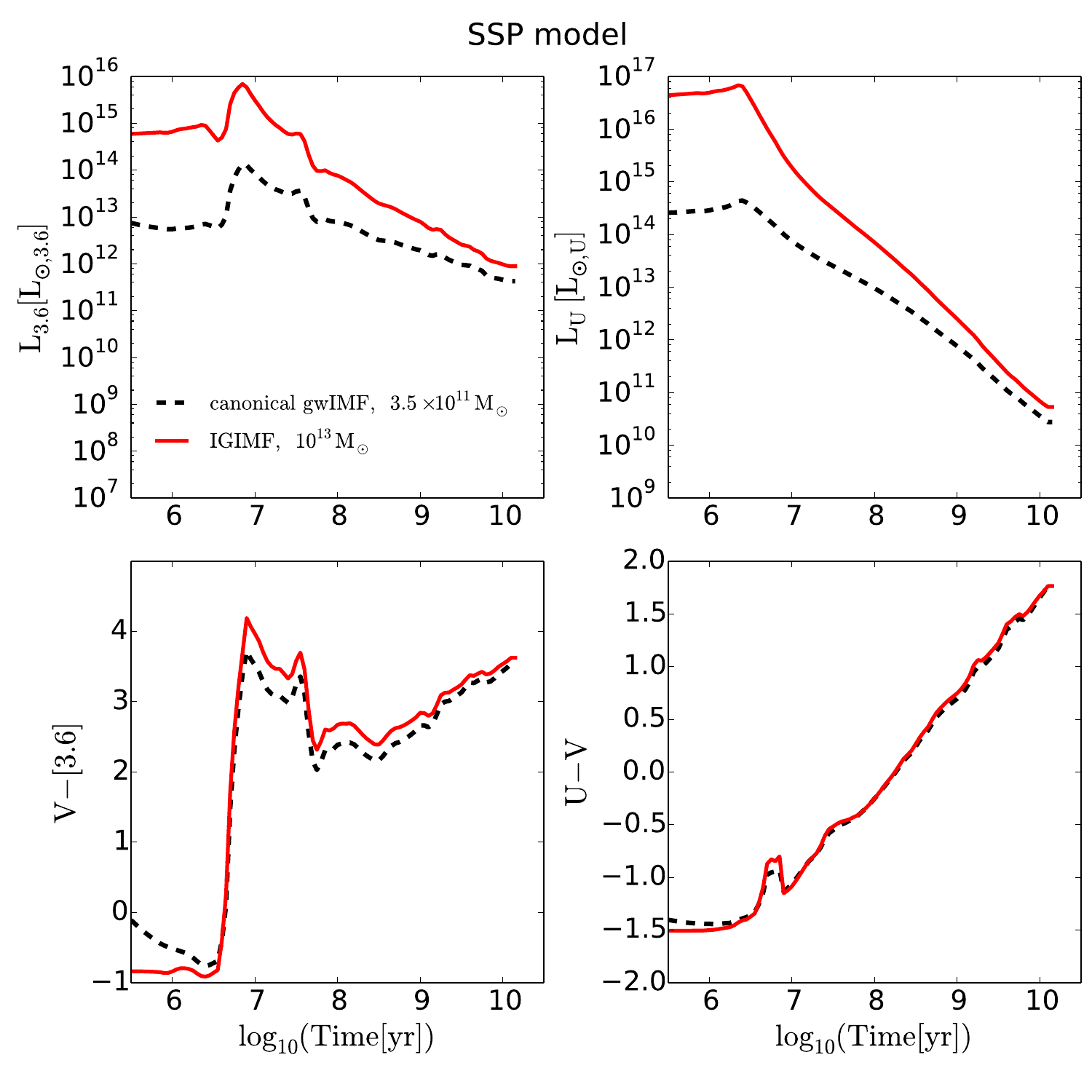}
\caption{Comparing the evolution of luminosity and color in different bands for two SSP galaxies in the framework of the invariant canonical gwIMF and the IGIMF. The horizontal axis refers to the time since the onset of star formation.}
\label{SSP-time}
\centering
\end{figure}

In Figure \ref{SSP-time} we plot the time evolution of color and luminosity for two modeled early-type galaxies with masses of $M_{\rm tot}=3.5\times10^{11}M_{\odot}$ and $10^{13} M_{\odot}$, constructed based on the invariant canonical gwIMF and the IGIMF, respectively. These galaxies are assumed to have been formed in a single burst of star formation 12.5 Gyr ago. As illustrated in Figure  \ref{SSP-time}, these two galaxies reach the same present-day $L_{[3.6]}$, $L_U$ and V-$[3.6]$, U-V color, while we know that their masses (or equivalently their $M/L_{[3.6]}$ ratios) differ by a factor of about 30. This is because the IGIMF theory leads massive galaxies with rapidly declining SFR to be dominated by dark remnants and have larger values of the $M/L_{[3.6]}$ ratios than what we expect from the invariant canonical gwIMF. This difference becomes milder by adopting a constant or slowly declining SFR. 

\begin{figure}
\includegraphics[width=9cm, height=10cm]{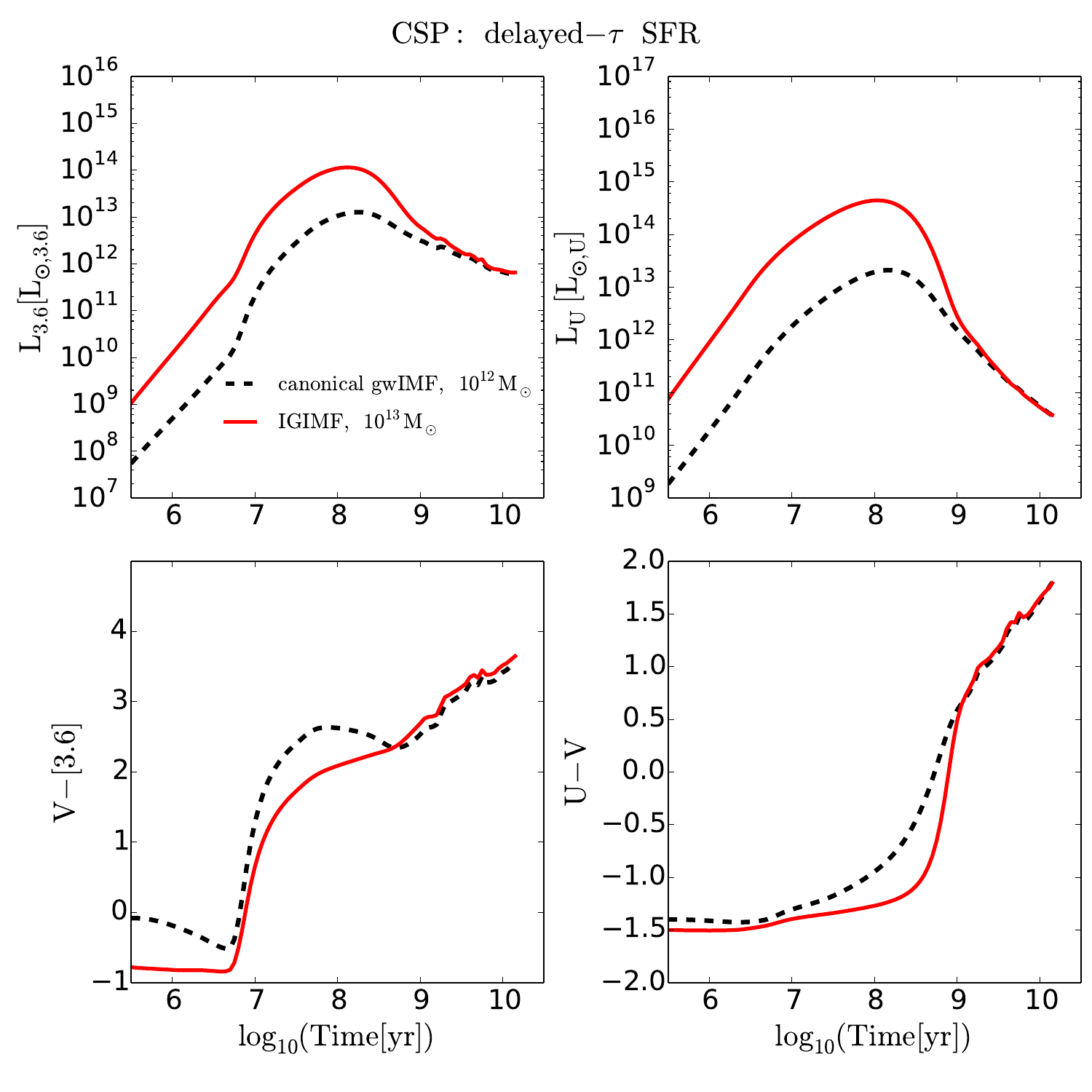}
\caption{Same as Figure \ref{SSP-time} but for two CSP galaxies with the delayed-$\tau$ SFR model with $\tau=$ 0.1 Gyr}
\label{Delay-time}
\centering
\end{figure}

\begin{figure}
\includegraphics[width=9cm, height=10cm]{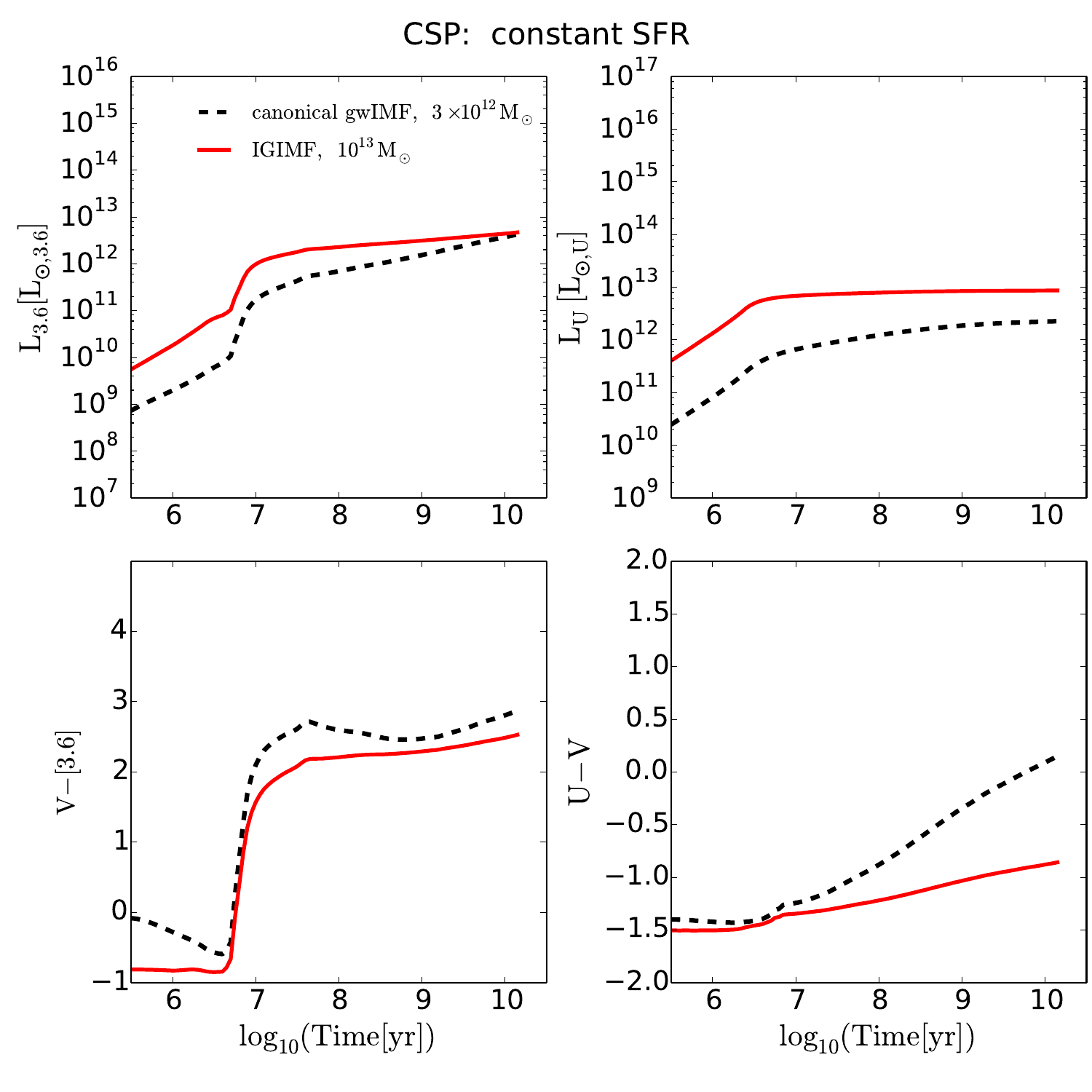}
\caption{Same as Figure \ref{Delay-time} but for two CSP galaxies with constant SFRs. }
\label{constant-time}
\centering
\end{figure}

Assuming a delayed-$\tau$ SFH, we compare two galaxies with stellar masses of  $M_{\rm tot}=10^{12}M_{\odot}$ and $10^{13} M_{\odot}$ in Figure \ref{Delay-time}. As can be seen, these galaxies have the same present-day luminosities in the [3.6] and U bands, as well as the same colors, making them observationally indistinguishable. By increasing the e-folding time scale, $\tau$, stars form in a larger time interval, resulting in less discrepancy in the masses of the galaxies constructed using the invariant canonical gwIMF versus the IGIMF. Figure \ref{constant-time} shows similar results but assumes a constant SFH for two galaxies with masses of  $M_{\rm tot}=3 \times 10^{12} M_\odot$ and $10^{13} M_\odot$ within the invariant canonical gwIMF and the IGIMF context, respectively.  These galaxies have similar luminosities in the [3.6] band, but their emitted light in the U band is noticeably different.  The significant difference in their U-V color can be used to differentiate between these two galaxies and break the degeneracy to determine their gwIMFs.

\begin{figure*}
\includegraphics[width=13cm, height=10cm]{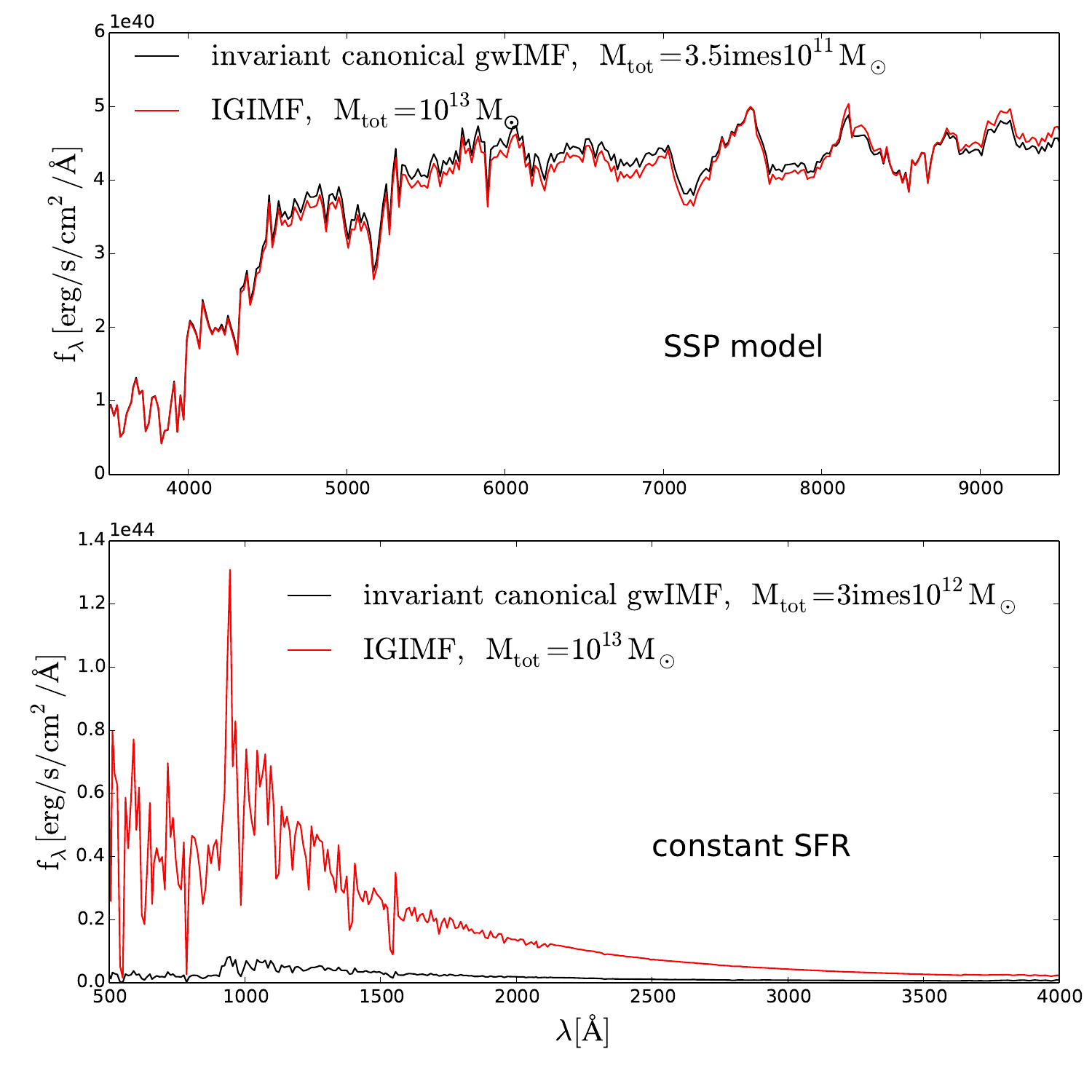}
\caption{SED of two galaxies with different masses and with the same luminosities in the [3.6] band constructed based on the invariant canonical gwIMF and the IGIMF for various SFHs at an age of 12.5 Gyr. Upper panel: SSP galaxies that have the same present-day  $L_{[3.6]}$ and $L_U$ as in Fig. \ref{SSP-time}. Lower panel: galaxies with constant SFR that have the same present-day  $L_{[3.6]}$ but different $L_U$ as in Fig. \ref{constant-time}. }
\label{SED}
\centering
\end{figure*}

We have compared the SEDs of two galaxies with different masses using various SFH models in Figure \ref{SED}. Our findings show that in the constant SFR model, the spectra of the two galaxies differ significantly at short wavelengths. However, the spectra of two galaxies in the SSP models (also in the delayed-$\tau$ model with the low value of $\tau$) are similar to each other in different wavelengths.

It should be noted that including binaries in SPS models can lead to changes in the resulting population's color, luminosity, and spectral properties. For example, in binary systems where mass transfer occurs, one star can accrete material from its companion, affecting its evolutionary path and final mass and hence potentially leading to the formation of a more massive remnant. Additionally, binary mergers can result in the formation of more massive remnants compared to single star evolution. Binary interactions can also influence the luminosity of stars. For example, in some cases, mass transfer can lead to enhanced stellar winds or increased accretion luminosity, affecting the overall brightness of the stellar population. Therefore, the presence of binary star systems has the overall effect of causing a population of stars to appear brighter and bluer at older ages than predicted by models focused only on single stars. Moreover, binary stars can impact the mass-to-light ratio of a stellar population through their influence on stellar masses and luminosities.  In later work we will consider the effect of binaries on these results (Zonoozi et al., in prep.)

\section{Conclusion}\label{Conc}

We have presented an SPS code (\texttt{SPS-VarIMF}) to calculate the stellar $M/L_X$ ratios in the context of a systematically varying stellar gwIMF with SFR and metallicity according to the IGIMF theory. The code provides the luminous stellar mass,  the fraction of dark remnants, luminosity in different bands, and the SED. Using \texttt{SPS-VarIMF} we calculate the time evolution of the $M/L$ ratios in the $[3.6]$-band and the V-band for different SFHs including the SSP model as well as constant and delayed-$\tau$ SFR models within both the invariant canonical gwIMF and the IGIMF contexts.  The main outcomes of our study can be summarized as follows:
\begin{itemize}
    \item We found that assuming the gwIMF is given by the invariant canonical IMF, increasing the metallicity leads to a decrease in the $M/L_{[3.6]}$ ratio. In the constant SFR case, the $M/L_{[3.6]}$ ratio changes from 0.9 to 0.5 (see Table 1 for more details).  In the context of the IGIMF, the difference between the $M/L_{[3.6]}$ ratio of massive galaxies and low-mass ones increases compared to the invariant canonical gwIMF such that for a wide range of galaxy masses, the present-day $M/L_{[3.6]}$ ratio varies by an order of magnitude depending on the underlying SFH of a galaxy.

     \item Our study suggests that early-type galaxies with different masses can exhibit similar observational properties, such as luminosity and color in various bands if they have different underlying gwIMFs. We show that if in reality, the stellar mass distribution follows the IGIMF theory, adopting the wrong gwIMF may lead to underestimating the mass of early-type galaxies even by an order of magnitude.

      \item  On the other hand, it is possible to identify the underlying gwIMF in late-type galaxies with a constant SFR or in recent starbursts by studying their colors and luminosities in different bands. These galaxies are dominated by young and luminous stars. Due to their high SFR, the gwIMF would be top-heavy according to the IGIMF theory which leads them to shine significantly in the $U-$band and have a completely different $U-V$ color than what is predicted based on the invariant canonical gwIMF.

      \item Finally, massive elliptical galaxies were $10^4$ times brighter when they were forming than today according to the IGIMF theory.
\end{itemize}

\section*{Data availability}
The data underlying this article are available in the article.

\section*{Acknowledgements}
AHZ acknowledges support through the AvH Foundation. PK thanks the DAAD Bonn-Prage Eastern European exchange program. HH thanks the staff at the HISKP and AIfA for their hospitality.

\appendix
\section{$M/L$ vs. color in different bands for galaxies with two different masses}

\begin{figure*}
	\centering
		
         \includegraphics[width=0.45\textwidth]{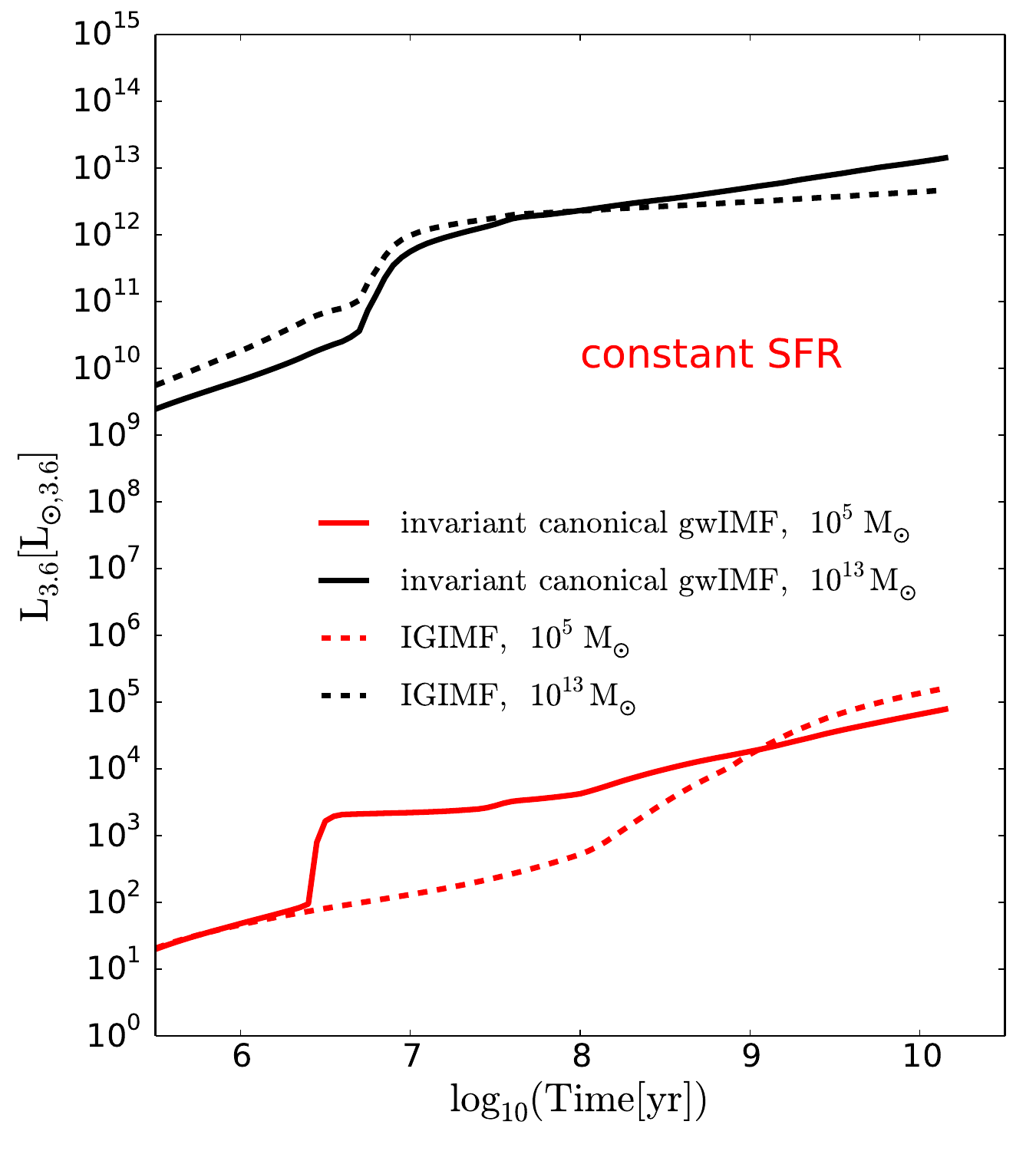}
         \includegraphics[width=0.45\textwidth]{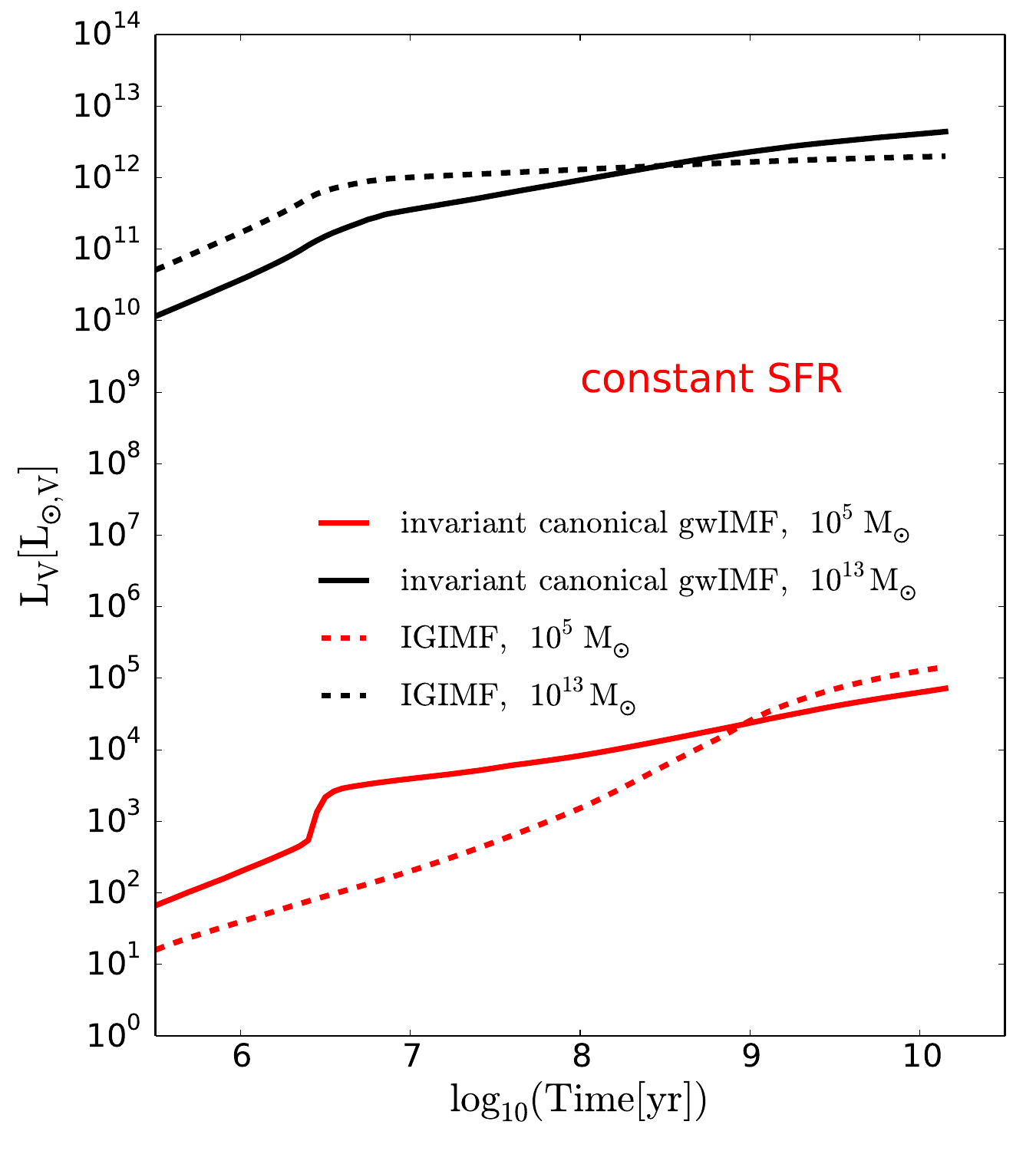}
         \includegraphics[width=0.45\textwidth]{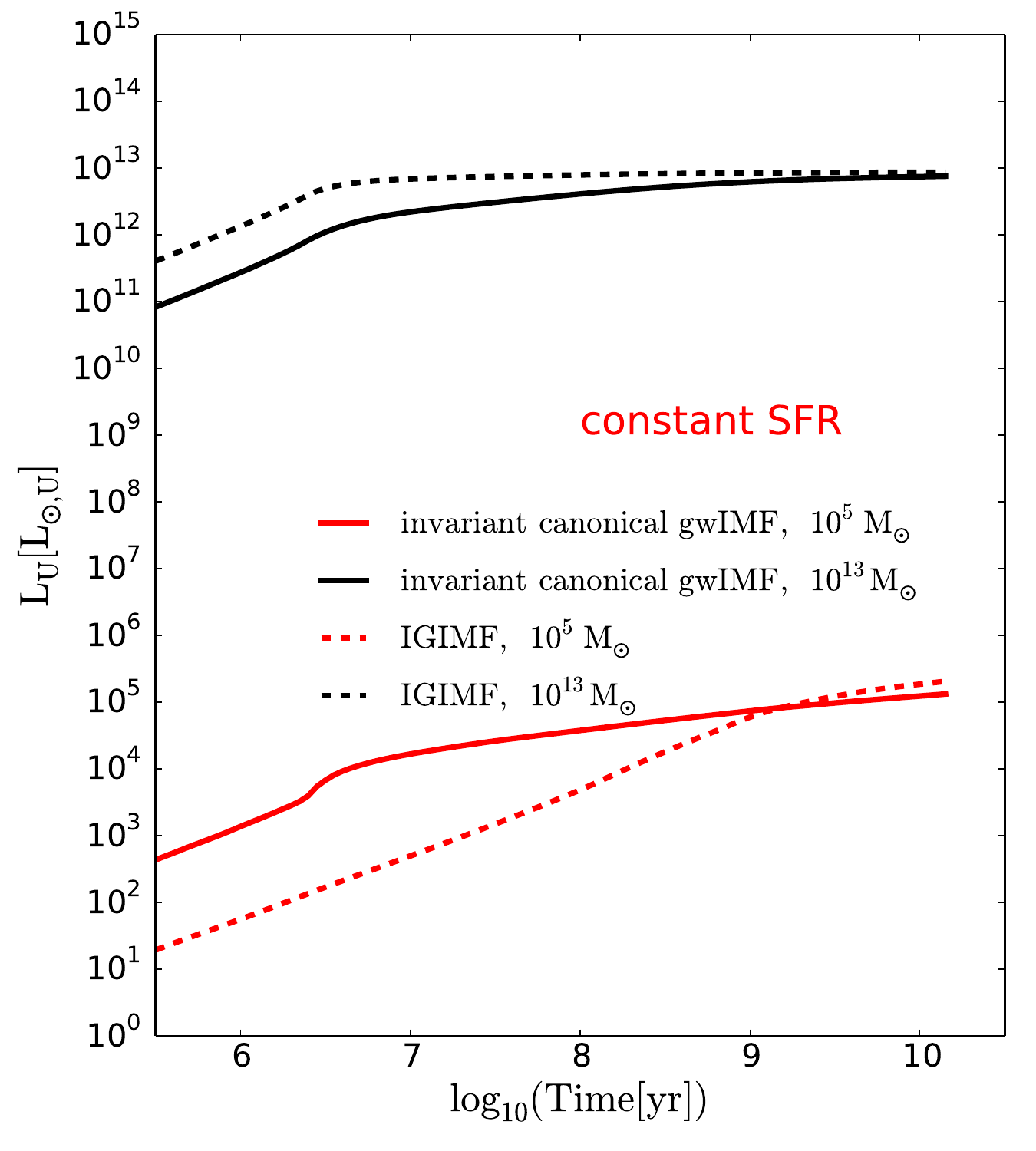}
         
	\caption{Same as Figure \ref{Bol1} but for three other bands ([3.6], V and U).}  
	\label{A1}
\end{figure*}

\begin{figure*}
	\centering
		
         \includegraphics[width=0.45\textwidth]{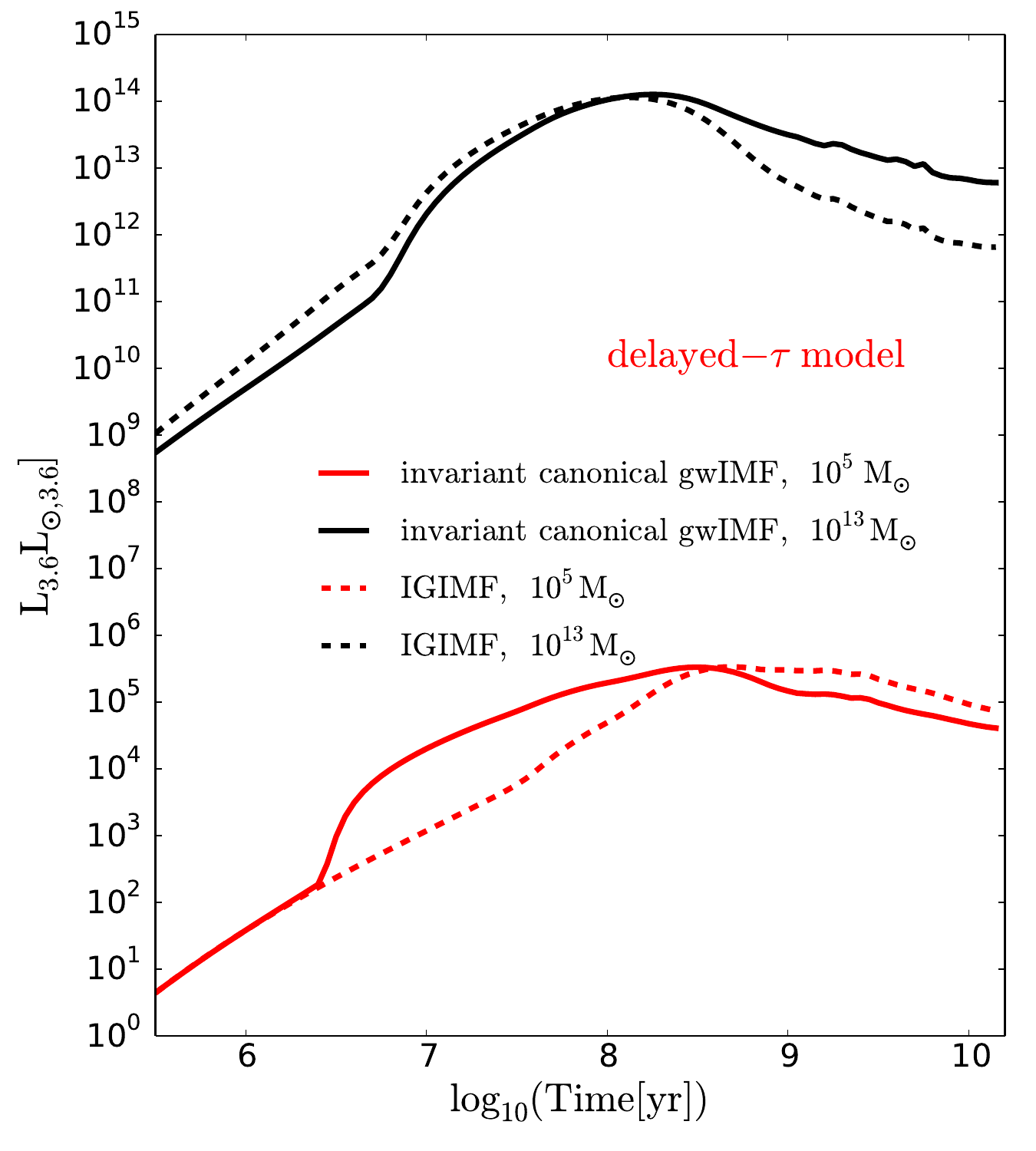}
         \includegraphics[width=0.45\textwidth]{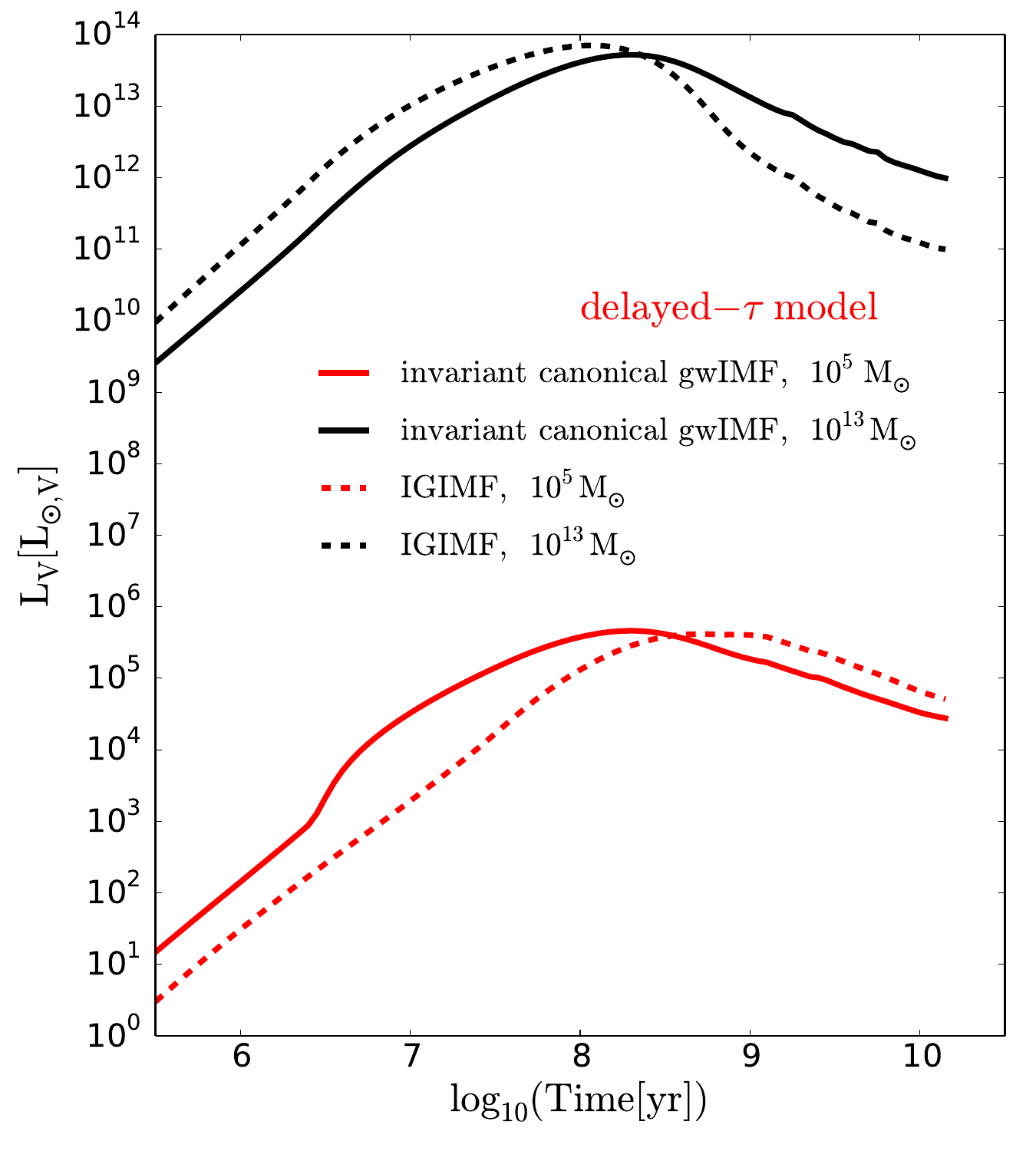}
         \includegraphics[width=0.45\textwidth]{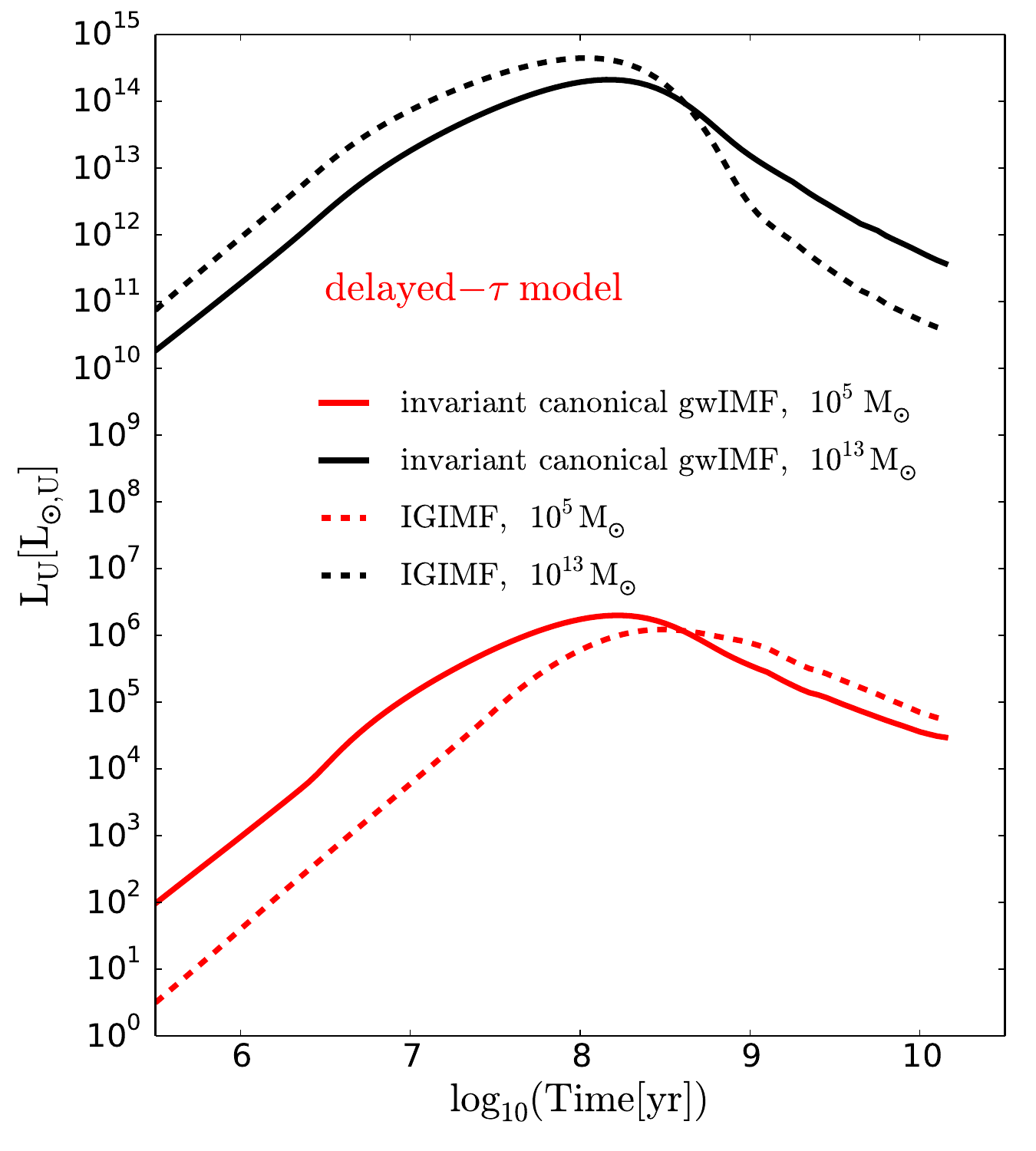}
        
	\caption{Same as Figure \ref{Bol2} but for three other bands ([3.6], V and U). }  
	\label{A2}
\end{figure*}

\begin{figure*}
	\centering
		
         \includegraphics[width=0.45\textwidth]{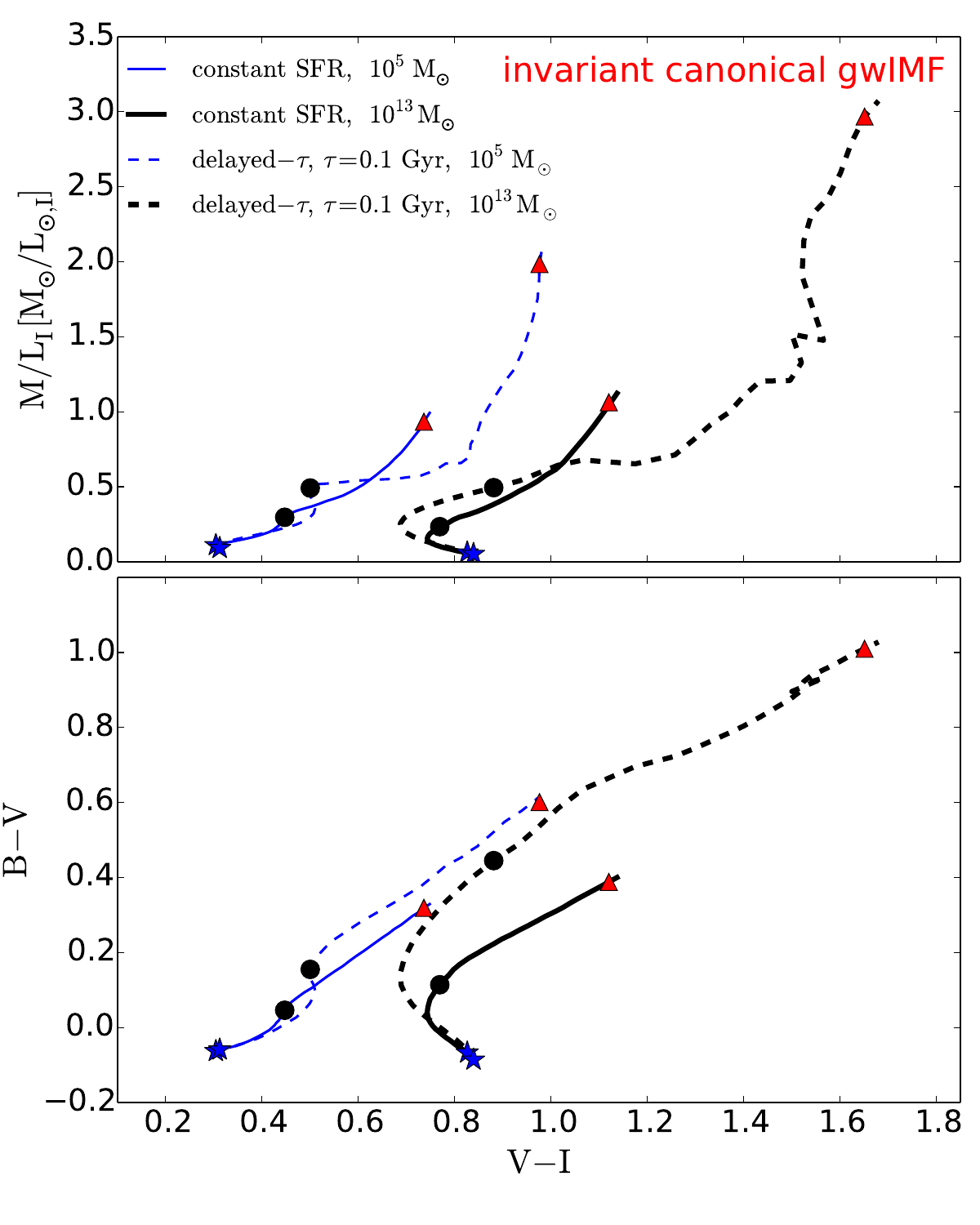}
         \includegraphics[width=0.45\textwidth]{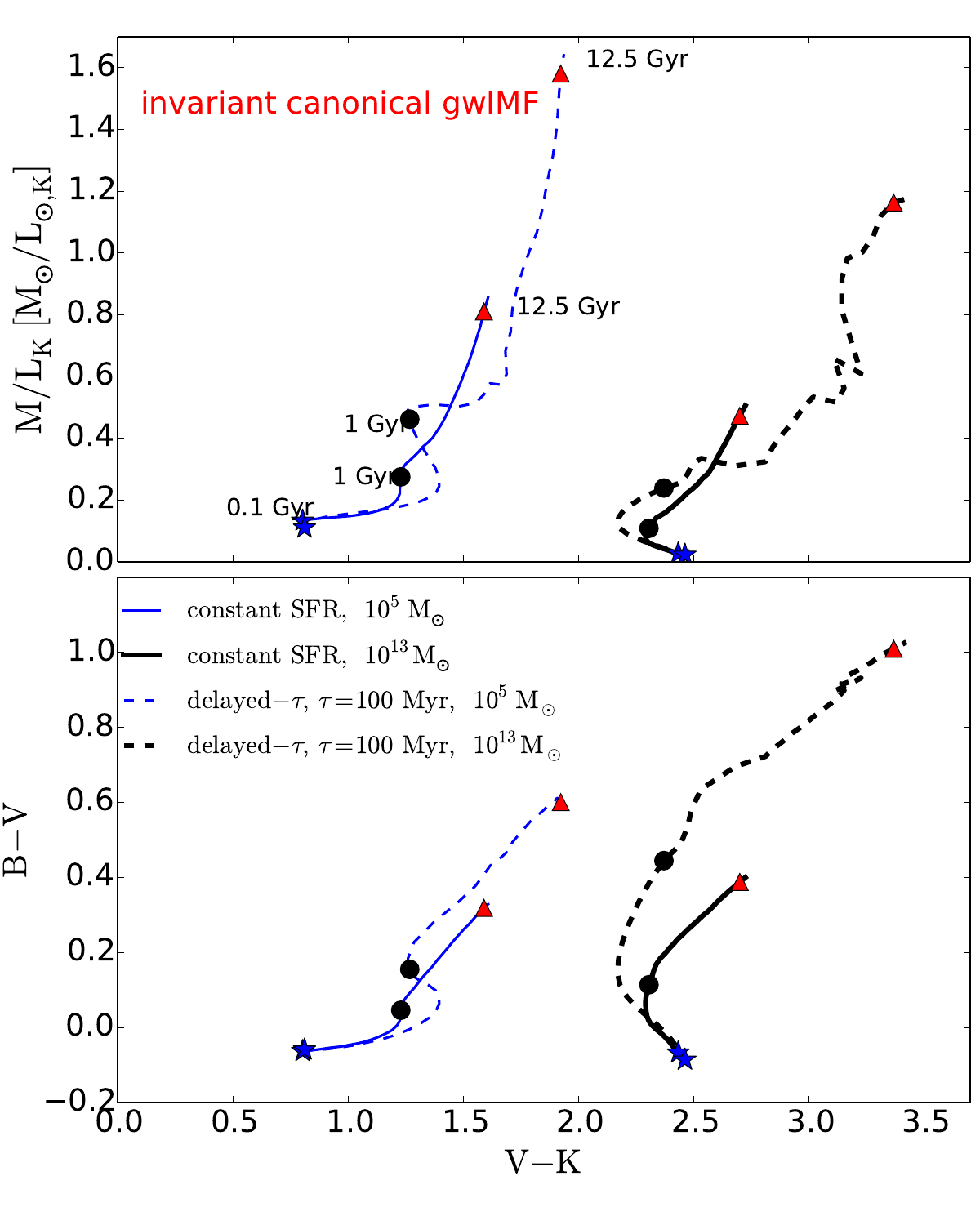}
	\caption{Same as Figure \ref{ML_T_Delay} but for two other bands (I and K) and colors (V-K and V-I).} 
	\label{A3}
\end{figure*}

\begin{figure*}
	\centering
		
         \includegraphics[width=0.45\textwidth]{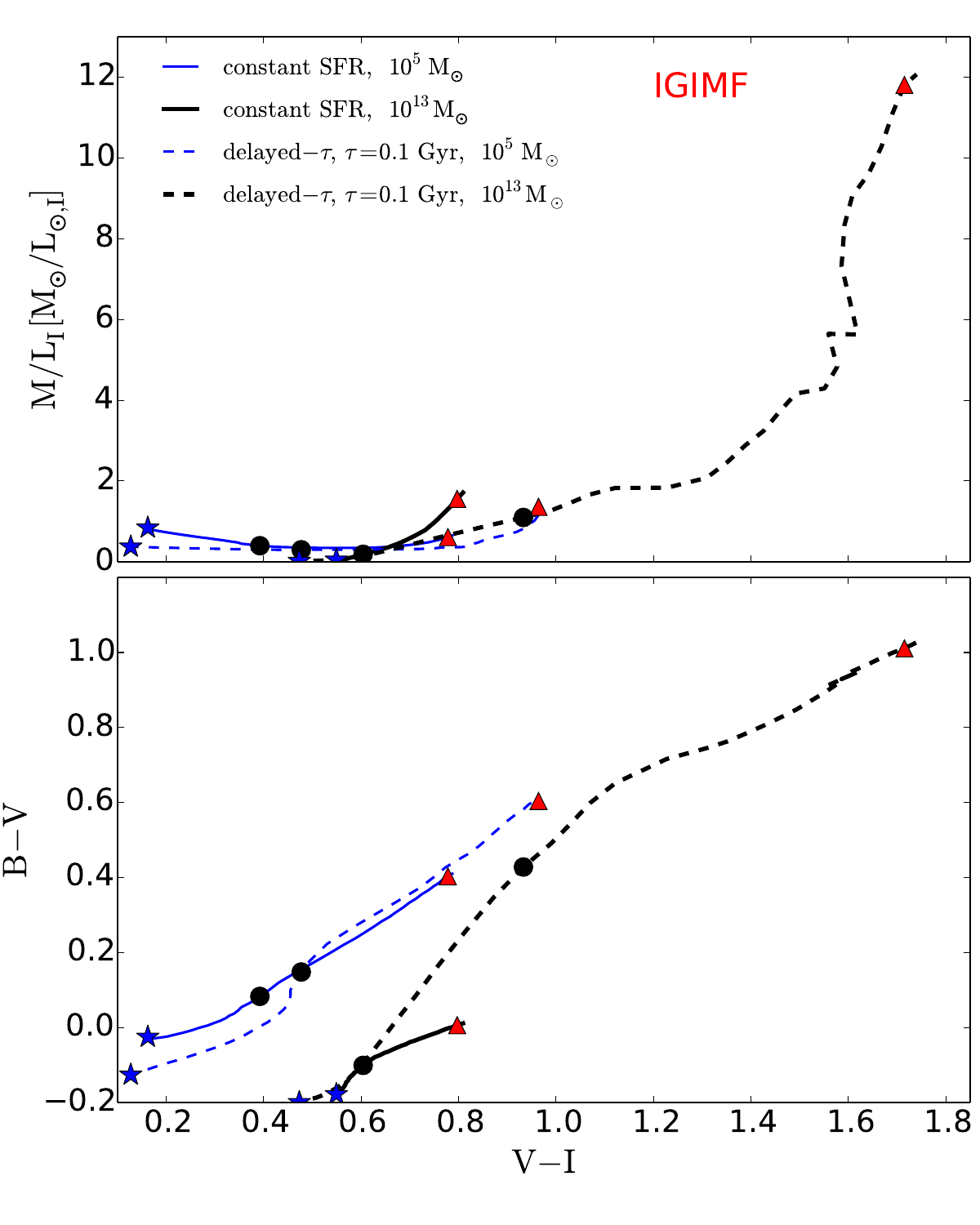}
         \includegraphics[width=0.45\textwidth]{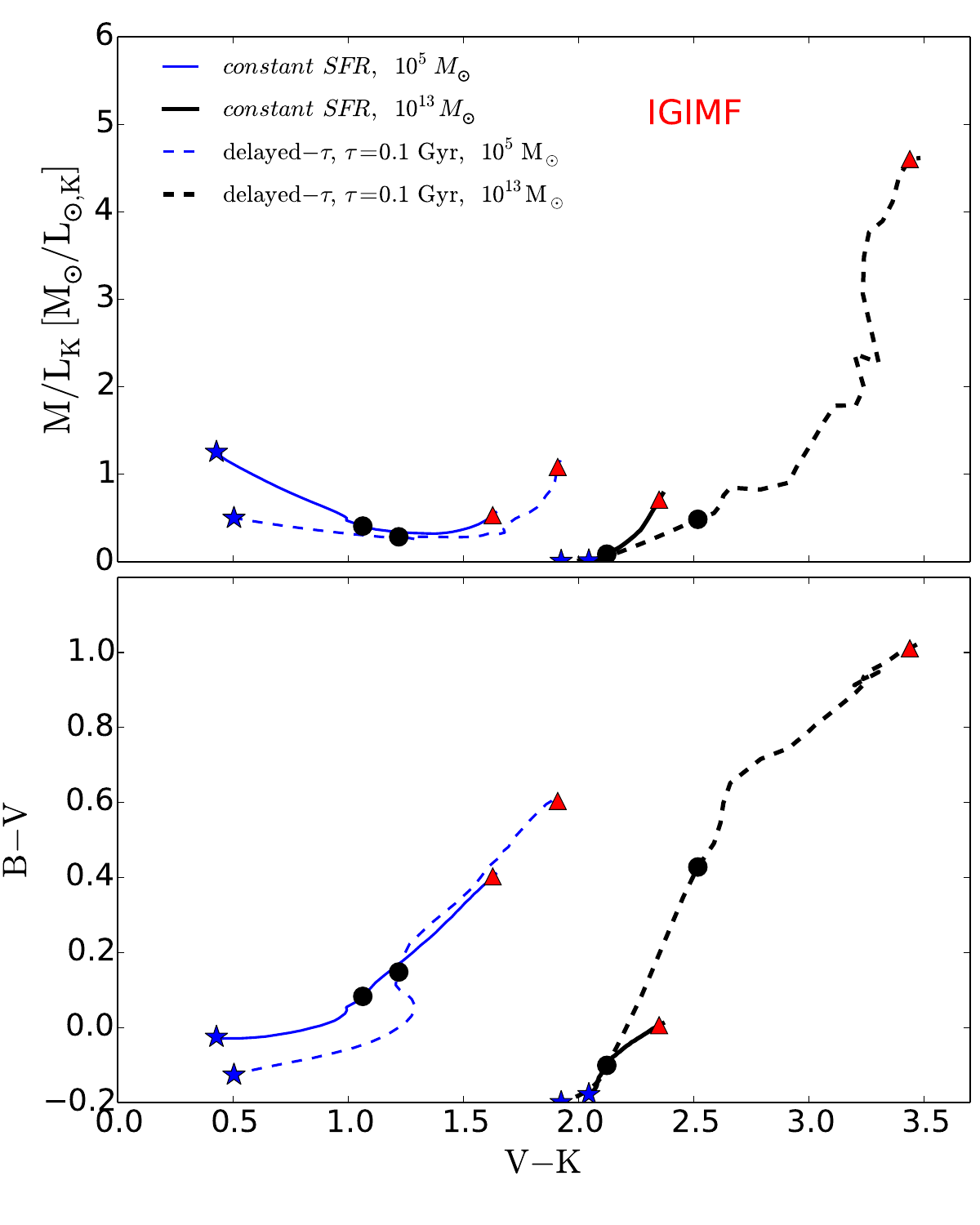}
	\caption{Same as Figure \ref{ML_T_Delay-IGIMF} but for two other bands (I and K) and colors (V-K and V-I).}
	\label{A4}
\end{figure*}

\bsp \label{lastpage} 
\begin{thebibliography}{99}





\bibitem[\protect\citeauthoryear{Asplund et al.}{2009}]{Asplund2009} Asplund M., Grevesse N., Sauval A.~J., Scott P., 2009, ARA\&A, 47, 481. 


\bibitem[\protect\citeauthoryear{Banerjee, Kroupa, \& Oh}{2012}]{Banerjee2012}
 Banerjee S., Kroupa P., Oh S., 2012, ApJ, 746, 15


\bibitem[\protect\citeauthoryear{Banik \& Zhao}{2022}]{Banik2022} Banik I., Zhao H., 2022, Symm, 14, 1331. 

\bibitem[\protect\citeauthoryear{Bastian, Covey \& Meyer}{2010}]{Bastian10} 
Bastian, N., Covey, K.~R., \& Meyer, M.~R.\ 2010, \araa, 48, 339

\bibitem[\protect\citeauthoryear{Bell \& de Jong}{2001}]{Bell2001} Bell E.~F., de Jong R.~S., 2001, ApJ, 550, 212. 

\bibitem[\protect\citeauthoryear{Bressan et al.}{2012}]{Bressan2012} Bressan A., Marigo P., Girardi L., Salasnich B., Dal Cero C., Rubele S., Nanni A., 2012, MNRAS, 427, 127. 


\bibitem[\protect\citeauthoryear{Bruzual \& Charlot}{2003}]{Bruzual03} 
Bruzual, G., \& Charlot, S. \ 2003, \mnras, 344, 1000

\bibitem[\protect\citeauthoryear{Calura et al.}{2010}]{Calura10}
 Calura, F., Recchi, S., Matteucci, F., \& Kroupa, P.\ 2010, \mnras, 406, 1985


\bibitem[\protect\citeauthoryear{Charlot \& Fall}{2000}]{Charlot2000} Charlot S., Fall S.~M., 2000, ApJ, 539, 718. 

\bibitem[\protect\citeauthoryear{Conroy, Gunn, \& White}{2009}]{Conroy2009} Conroy C., Gunn J.~E., White M., 2009, ApJ, 699, 486. 

\bibitem[\protect\citeauthoryear{Conroy et al.}{2017}]{Conroy17} 
Conroy, C., van Dokkum, P.~G., \& Villaume, A.\ 2017, \apj, 837, 166 


\bibitem[\protect\citeauthoryear{Cowie et al.}{1996}]{Cowie1996} Cowie L.~L., Songaila A., Hu E.~M., Cohen J.~G., 1996, AJ, 112, 839. 



\bibitem[\protect\citeauthoryear{Dabringhausen et al.}{2012}]{Dabringhausen12}
 Dabringhausen, J., Kroupa, P., Pflamm-Altenburg, J., \& Mieske, S.\ 2012, \apj, 747, 72

\bibitem[\protect\citeauthoryear{Dabringhausen \& Kroupa}{2023}]{Dabring2023} Dabringhausen J., Kroupa P., 2023, MNRAS, 526, 2301. 


\bibitem[\protect\citeauthoryear{Dinnbier, Kroupa, \& Anderson}{2022}]{Dinnbier2022} Dinnbier F., Kroupa P., Anderson R.~I., 2022, A\&A, 660, A61. 


\bibitem[\protect\citeauthoryear{Eappen et al.}{2022}]{Eappen2022} Eappen R., Kroupa P., Wittenburg N., Haslbauer M., Famaey B., 2022, MNRAS, 516, 1081. 

\bibitem[\protect\citeauthoryear{Eldridge et al.}{2017}]{Eldridge2017} Eldridge J.~J., Stanway E.~R., Xiao L., McClelland L.~A.~S., Taylor G., Ng M., Greis S.~M.~L., et al., 2017, PASA, 34, e058. 



\bibitem[\protect\citeauthoryear{Eldridge \& Stanway}{2022}]{Eldridge2022} Eldridge J.~J., Stanway E.~R., 2022, ARA\&A, 60, 455. 

\bibitem[\protect\citeauthoryear{Famaey \& McGaugh}{2012}]{Famaey2012} Famaey B., McGaugh S.~S., 2012, LRR, 15, 10. 


\bibitem[\protect\citeauthoryear{Fontanot et al.}{2017}]{Fontanot2017}
Fontanot F., De Lucia G., Hirschmann M., Bruzual G., Charlot S., Zibetti S., 2017, MNRAS, 464, 3812


\bibitem[\protect\citeauthoryear{Gallazzi et al.}{2005}]{Gallazzi2005} Gallazzi A., Charlot S., Brinchmann J., White S.~D.~M., Tremonti C.~A., 2005, MNRAS, 362, 41. 

\bibitem[\protect\citeauthoryear{Geha et al.}{2013}]{Geha13}
 Geha, M., Brown, T.~M., Tumlinson, J., et al.\ 2013, \apj, 771, 29 

\bibitem[\protect\citeauthoryear{Gennaro et al.}{2018}]{Gennaro18} 
Gennaro, M., Tchernyshyov, K., Brown, T.~M., et al.\ 2018, \apj, 855, 20 


\bibitem[\protect\citeauthoryear{Gjergo et al.}{2023}]{Gjergo2023} Gjergo E., Sorokin A.~G., Ruth A., Spitoni E., Matteucci F., Fan X., Liang J., et al., 2023, ApJS, 264, 44. 

\bibitem[\protect\citeauthoryear{Gunawardhana et al.}{2011}]{Gunawardhana11}
 Gunawardhana, M.~L.~P., Hopkins, A.~M., Sharp, R.~G., et al.\ 2011, \mnras, 415, 1647


\bibitem[\protect\citeauthoryear{Habergham et al.}{2010}]{Habergham10}
 Habergham, S.~M., Anderson, J.~P., \& James, P.~A.\ 2010, \apj, 717, 342

 \bibitem[\protect\citeauthoryear{Haslbauer, Kroupa, \& Jerabkova}{2023}]{Haslbauer2023} Haslbauer M., Kroupa P., Jerabkova T., 2023, MNRAS, 524, 3252. 


\bibitem[\protect\citeauthoryear{Haslbauer et al.}{2024}]{Haslbauer2024} Haslbauer M., Yan Z., Jerabkova T., Gjergo E., Kroupa P., Hasani Zonoozi A., 2024, A\&A, 689, A221. 
 

\bibitem[\protect\citeauthoryear{Heger et al.}{2003}]{Heger2003} Heger A., Fryer C.~L., Woosley S.~E., Langer N., Hartmann D.~H., 2003, ApJ, 591, 288. 

\bibitem[\protect\citeauthoryear{Hopkins et al.}{2013}]{Hopkins13}
 Hopkins, A.~M., Driver, S.~P., Brough, S., et al.\ 2013, \mnras, 430, 2047

 \bibitem[\protect\citeauthoryear{Hopkins}{2018}]{Hopkins18} Hopkins A.~M., 2018, PASA, 35, e039. 

\bibitem[\protect\citeauthoryear{Hoversten \& Glazebrook}{2008}]{Hoversten08}
 Hoversten, E.~A., \& Glazebrook, K.\ 2008, \apj, 675, 163-187

\bibitem[\protect\citeauthoryear{Je{\v{r}}{\'a}bkov{\'a} et al.}{2018}]{Jerabkova2018} Je{\v{r}}{\'a}bkov{\'a} T., Hasani Zonoozi A., Kroupa P., Beccari G., Yan Z., Vazdekis A., Zhang Z.-Y., 2018, A\&A, 620, A39. 
 
\bibitem[\protect\citeauthoryear{Joncour et al.}{2018}]{Joncour18} Joncour I., Duch{\^e}ne G., Moraux E., Motte F., 2018, A\&A, 620, A27. 
 

\bibitem[\protect\citeauthoryear{Kalari et al.}{2018}]{Kalari18}
  Kalari, V.~M., Carraro, G., Evans, C.~J., \& Rubio, M., 2018, ApJ, 857, 132



\bibitem[\protect\citeauthoryear{Kirk \& Myers}{2012}]{Kirk12}
 Kirk, H., \& Myers, P.~C.\ 2012, \apj, 745, 131

 

\bibitem[\protect\citeauthoryear{K{\"o}ppen, Weidner, \& Kroupa}{2007}]{Kopen2007}
 K{\"o}ppen J., Weidner C., Kroupa P., 2007, MNRAS, 375, 673


\bibitem[\protect\citeauthoryear{Kroupa et al.}{1993}]{Kroupa93} 
Kroupa, P., Tout, C.~A., \& Gilmore, G.\ 1993, \mnras, 262, 545

\bibitem[\protect\citeauthoryear{Kroupa}{1995a}]{Kroupa1995a} Kroupa P., 1995, MNRAS, 277, 1491. 

\bibitem[\protect\citeauthoryear{Kroupa}{1995b}]{Kroupa1995b}Kroupa P., 1995, MNRAS, 277, 1507. 



\bibitem[\protect\citeauthoryear{Kroupa}{2001}]{Kroupa01} 
 Kroupa, P.\ 2001, \mnras, 322, 231


\bibitem[\protect\citeauthoryear{Kroupa}{2002}]{Kroupa02}
 Kroupa, P.\ 2002, Science, 295, 82
 

\bibitem[\protect\citeauthoryear{Kroupa \& Bouvier}{2003}]{KroupaBovier03}
 Kroupa, P., \& Bouvier, J.\ 2003, \mnras, 346, 369



\bibitem[\protect\citeauthoryear{Kroupa \& Weidner}{2003}]{Kroupa03}
 Kroupa, P., \& Weidner, C.\ 2003, \apj, 598, 1076

\bibitem[\protect\citeauthoryear{Kroupa}{2005}]{Kroupa05}
Kroupa P., 2005, in Turon C., OFlaherty K. S., Perryman M. A. C., eds, ESA
SP-576: The Three-Dimensional Universe with Gaia. ESA, Noordwijk,
p. 629

\bibitem[\protect\citeauthoryear{Kroupa et al.}{2013}]{Kroupa13}
 Kroupa, P., Weidner, C., Pflamm-Altenburg, J., et al.\ 2013, Planets, Stars and Stellar Systems.~Volume 5: Galactic Structure and Stellar Populations, 5, 115

 \bibitem[\protect\citeauthoryear{Kroupa et al.}{2020}]{Kroupa2020} Kroupa P., Haslbauer M., Banik I., Nagesh S.~T., Pflamm-Altenburg J., 2020, MNRAS, 497, 37. 



 \bibitem[\protect\citeauthoryear{Kroupa et al.}{2023}]{Kroupa2023b} Kroupa P., Gjergo E., Asencio E., Haslbauer M., Pflamm-Altenburg J., Wittenburg N., Samaras N., et al., 2023, arXiv, arXiv:2309.11552. 

 \bibitem[\protect\citeauthoryear{Kroupa et al.}{2024}]{Kroupa2024} Kroupa P., Gjergo E., Jerabkova T., Yan Z., 2024, arXiv, arXiv:2410.07311. 


\bibitem[\protect\citeauthoryear{Lada \& Lada}{2003}]{Lada03}
 Lada C. J., Lada E. A., 2003, ARA\&A, 41, 57

\bibitem[\protect\citeauthoryear{Lee et al.}{2009}]{Lee09}
 Lee, J.~C., Gil de Paz, A., Tremonti, C., et al.\ 2009, \apj, 706, 599-613




 \bibitem[\protect\citeauthoryear{Lejeune, Cuisinier, \& Buser}{1997}]{Lejeune1997} Lejeune T., Cuisinier F., Buser R., 1997, A\&AS, 125, 229. 

\bibitem[\protect\citeauthoryear{Lejeune, Cuisinier, \& Buser}{1998}]{Lejeune1998} Lejeune T., Cuisinier F., Buser R., 1998, A\&AS, 130, 65. 

 
\bibitem[\protect\citeauthoryear{Mansfield \& Kroupa}{2023}]{Mansfield2023} Mansfield S., Kroupa P., 2023, MNRAS, 525, 6005. 


\bibitem[\protect\citeauthoryear{Marigo}{2001}]{Marigo2001} Marigo P., 2001, A\&A, 370, 194. 

 \bibitem[\protect\citeauthoryear{Marigo et al.}{2017}]{Marigo2017} Marigo P., Girardi L., Bressan A., Rosenfield P., Aringer B., Chen Y., Dussin M., et al., 2017, ApJ, 835, 77. 

\bibitem[\protect\citeauthoryear{Marks \& Kroupa}{2012}]{Marks12a} 

\bibitem[\protect\citeauthoryear{Marks et al.}{2012}]{Marks12}
 Marks, M., Kroupa, P., Dabringhausen, J., \& Pawlowski, M.~S.\ 2012, \mnras, 422, 2246


\bibitem[\protect\citeauthoryear{Matteucci \& Brocato}{1990}]{Matteucci90}
 Matteucci, F., \& Brocato, E.\ 1990, \apj, 365, 539

 \bibitem[\protect\citeauthoryear{McDermid et al.}{2015}]{McDermid2015} McDermid R.~M., Alatalo K., Blitz L., Bournaud F., Bureau M., Cappellari M., Crocker A.~F., et al., 2015, MNRAS, 448, 3484. 

\bibitem[\protect\citeauthoryear{Megeath et al.}{2016}]{Megeath16}
 Megeath, S.~T., Gutermuth, R., Muzerolle, J., et al.\ 2016, \aj, 151, 5

\bibitem[\protect\citeauthoryear{Meurer et al.}{2009}]{Meurer09}
 Meurer, G.~R., Wong, O.~I., Kim, J.~H., et al.\ 2009, \apj, 695, 765

\bibitem[\protect\citeauthoryear{Milgrom}{1983}]{Milgrom1983} Milgrom M., 1983, ApJ, 270, 365. 
 
\bibitem[\protect\citeauthoryear{Offner et al.}{2014}]{Offner14} 
 Offner, S.~S.~R., Clark, P.~C., Hennebelle, P., et al.\ 2014, Protostars and Planets VI, 53


\bibitem[\protect\citeauthoryear{Pflamm-Altenburg \& Kroupa}{2008}]{Pflamm08} 
Pflamm-Altenburg, J., \& Kroupa, P.\ 2008, \nat, 455, 641


\bibitem[\protect\citeauthoryear{Pflamm-Altenburg et al.}{2011}]{Pflamm11} 
Pflamm-Altenburg, J., Weidner, C., \& Kroupa, P.\ 2011, UP2010: Have Observations Revealed a Variable Upper End of the Initial Mass Function?, 440, 269

\bibitem[\protect\citeauthoryear{Recchi et al.}{2009}]{Recchi09} 
 Recchi, S., Calura, F., \& Kroupa, P.\ 2009, \aap, 499, 711

 \bibitem[\protect\citeauthoryear{Recchi \& Kroupa}{2015}]{Recchi2015} Recchi S., Kroupa P., 2015, MNRAS, 446, 4168. 

 \bibitem[\protect\citeauthoryear{Renzini \& Ciotti}{1993}]{Renzini1993} Renzini A., Ciotti L., 1993, ApJL, 416, L49. 

\bibitem[\protect\citeauthoryear{Scalo}{1986}]{Scalo86}
 Scalo, J.~M.\ 1986, \fcp, 11, 1


\bibitem[\protect\citeauthoryear{Scalo}{1998}]{Scalo98}
Scalo, J.\ 1998, The Stellar Initial Mass Function (38th Herstmonceux Conference), 142, 201  

\bibitem[\protect\citeauthoryear{Schmidt}{1959}]{Schmidt1959} Schmidt M., 1959, ApJ, 129, 243. 

\bibitem[\protect\citeauthoryear{Schneider et al.}{2018}]{Schneider18} 
Schneider, F.~R.~N., Sana, H., Evans, C.~J., et al.\ 2018, Science, 359, 69

\bibitem[\protect\citeauthoryear{Schulz et al.}{2015}]{Schulz15}
Schulz, C., Pflamm-Altenburg, J., \& Kroupa, P.\ 2015, \aap, 582, A93

\bibitem[\protect\citeauthoryear{Schombert}{2016}]{Schombert2016} Schombert J.~M., 2016, AJ, 152, 214. 

\bibitem[\protect\citeauthoryear{Schombert, McGaugh, \& Lelli}{2019}]{Schombert2019} Schombert J., McGaugh S., Lelli F., 2019, MNRAS, 483, 1496. 

\bibitem[\protect\citeauthoryear{Schombert, McGaugh, \& Lelli}{2020}]{Schombert2020} Schombert J., McGaugh S., Lelli F., 2020, AJ, 160, 71. 

\bibitem[\protect\citeauthoryear{Spera, Mapelli, \& Bressan}{2015}]{Spera2015} Spera M., Mapelli M., Bressan A., 2015, MNRAS, 451, 4086. 


\bibitem[\protect\citeauthoryear{Spiniello et al.}{2015}]{Spiniello15}
Spiniello, C., Trager, S.~C., \& Koopmans, L.~V.~E.\ 2015, \apj, 803, 87 


\bibitem[\protect\citeauthoryear{Stanway \& Eldridge}{2019}]{Stanway2019} Stanway E.~R., Eldridge J.~J., 2019, A\&A, 621, A105. 

\bibitem[\protect\citeauthoryear{Stanway et al.}{2020}]{Stanway2020} Stanway E.~R., Chrimes A.~A., Eldridge J.~J., Stevance H.~F., 2020, MNRAS, 495, 4605. 

\bibitem[\protect\citeauthoryear{Stanway \& Eldridge}{2023}]{Stanway2023} Stanway E.~R., Eldridge J.~J., 2023, MNRAS, 522, 4430. 


\bibitem[\protect\citeauthoryear{Tinsley}{1980}]{Tinsley80} 
 Tinsley, B.~M.\ 1980, \fcp, 5, 287


\bibitem[\protect\citeauthoryear{Thomas et al.}{2005}]{Thomas2005} Thomas D., Maraston C., Bender R., Mendes de Oliveira C., 2005, ApJ, 621, 673. 

\bibitem[\protect\citeauthoryear{{\'U}beda et al.}{2007}]{Ubeda07} 
{\'U}beda, L., Ma{\'{\i}}z-Apell{\'a}niz, J., \& MacKenty, J.~W.\ 2007, \aj, 133, 932
  
\bibitem[\protect\citeauthoryear{van Dokkum \& Conroy}{2011}]{vanDokkum11}
 van Dokkum, P.~G., \& Conroy, C.\ 2011, \apjl, 735, L13

\bibitem[\protect\citeauthoryear{Vazdekis et al.}{2003}]{Vazdekis03}
Vazdekis, A., Cenarro, A.~J., Gorgas, J., Cardiel, N., \& Peletier, R.~F.\ 2003, \mnras, 340, 1317

\bibitem[\protect\citeauthoryear{Vazdekis et al.}{2012}]{Vazdekis2012} Vazdekis A., Ricciardelli E., Cenarro A.~J., Rivero-Gonz{\'a}lez J.~G., D{\'\i}az-Garc{\'\i}a L.~A., Falc{\'o}n-Barroso J., 2012, MNRAS, 424, 157. 
 
\bibitem[\protect\citeauthoryear{Watts et al.}{2018}]{Watts18}
 Watts, A.~B., Meurer, G.~R., Lagos, C.~D.~P., et al.\ 2018, \mnras, 477, 5554 
 

\bibitem[\protect\citeauthoryear{Weidner \& Kroupa}{2006}]{Weidner06} 
Weidner, C. \& Kroupa, P. 2006, MNRAS, 365, 1333, 2


\bibitem[\protect\citeauthoryear{Weidner et al.}{2013b}]{Weidner13b}
 Weidner, C., Kroupa, P., Pflamm-Altenburg, J., \& Vazdekis, A. 2013b, \mnras, 436, 3309

 \bibitem[\protect\citeauthoryear{Westera et al.}{2002}]{Westera2002} Westera P., Lejeune T., Buser R., Cuisinier F., Bruzual G., 2002, A\&A, 381, 524.  

\bibitem[\protect\citeauthoryear{Yan, Je{\v{r}}{\'a}bkov{\'a}, \& Kroupa}{2017}] {Yan2017}
 Yan Z., Jerabkova T., Kroupa P., 2017, A\&A, 607, A126
 
\bibitem[\protect\citeauthoryear{Yan, Je{\v{r}}{\'a}bkov{\'a}, \& Kroupa}{2020}] {Yan2020}
 Yan, Z., Jerabkova, T., \& Kroupa, P. 2020, A\&A, 637, A68

 \bibitem[\protect\citeauthoryear{Yan, Je{\v{r}}{\'a}bkov{\'a}, \& Kroupa}{2021}]{Yan2021} Yan Z., Je{\v{r}}{\'a}bkov{\'a} T., Kroupa P., 2021, A\&A, 655, A19. 

\bibitem[\protect\citeauthoryear{Yan, Je{\v{r}}{\'a}bkov{\'a}, \& Kroupa}{2023}]{Yan2023} Yan Z., Jerabkova T., Kroupa P., 2023, A\&A, 670, A151. 

\bibitem[\protect\citeauthoryear{Yan et al.}{2024}]{Yan2024} Yan Z., Li J., Kroupa P., Jerabkova T., Gjergo E., Zhang Z.-Y., 2024, ApJ, 969, 95. 


\bibitem[\protect\citeauthoryear{Yang et al.}{2024}]{Yan2024b} Yang Y., Liu C., Yang M., Zheng Y., Tian H., 2024, ApJ, 977, 11. 


\bibitem[\protect\citeauthoryear{Yasui et al.}{2023}]{Yasui2023} Yasui C., Kobayashi N., Saito M., Izumi N., Ikeda Y., 2023, ApJ, 943, 137. 

\bibitem[\protect\citeauthoryear{Zhang \& Fall}{1999}]{Zhang99} 
Zhang, Q., \& Fall, S.~M.\ 1999, \apjl, 527, L81
 
 
\bibitem[\protect\citeauthoryear{Zhang et al.}{2018}]{Zhang18}
 Zhang, Z.-Y., Romano, D., Ivison, R.~J., Papadopoulos, P.~P., \& Matteucci, F.\ 2018, Natur, 558, 260  

 \bibitem[\protect\citeauthoryear{Zinnkann, Wirth, \& Kroupa}{2024}]{Zinnkann2024} Zinnkann M., Wirth H., Kroupa P., 2024, arXiv, arXiv:2402.09405. 


\end{thebibliography}
\end{document}